\documentclass[12pt]{article}
\usepackage{epsfig,amsfonts,amsthm}
\usepackage{amsmath,amssymb}
\usepackage{array}
\usepackage{cite}
\usepackage{mathrsfs}
\usepackage[title]{appendix}
\usepackage{cancel}
\usepackage{slashed}
\usepackage{hyperref}
\usepackage{mathtools}

\newcommand{\Et}{\cancel{E}_{T}}
\newcommand{\be}{\begin{equation}}
\newcommand{\ee}{\end{equation}}
\newcommand{\bea}{\begin{eqnarray}}
\newcommand{\eea}{\end{eqnarray}}

\newcommand{\doublet}[2]{ \left( \begin{array}{c}#1 \\ #2 \end{array}\right) }

\newcommand{\GeV}{{\ensuremath\rm \; GeV}}

\def\lsim{\mathrel{\rlap{\lower4pt\hbox{\hskip1pt$\sim$}}
    \raise1pt\hbox{$<$}}}         
\def\gsim{\mathrel{\rlap{\lower4pt\hbox{\hskip1pt$\sim$}}
    \raise1pt\hbox{$>$}}}         

\newcommand{\bft}{\begin{footnotesize}}
\newcommand{\eft}{\end{footnotesize}}
\newcommand{\bts}{\begin{tiny}}
\newcommand{\ets}{\end{tiny}}

\newcommand{\bt}{\begin{tabular}}
\newcommand{\et}{\end{tabular}}

\usepackage{color}
\definecolor{grey}{cmyk}{0,0,0,0.75}
\definecolor{tangerine}{cmyk}{0,0.5,1,0}
\definecolor{darkgreen}{cmyk}{1,0,1,0.23}

\definecolor{Red}{rgb}{0.9,0.1,0.1}
\definecolor{Blue}{rgb}{0.1,0.4,0.8}
\definecolor{Green}{rgb}{0.1,0.8,0.4}
\definecolor{BlueGreen}{rgb}{0,0.8,0.5}
\definecolor{yellow}{cmyk}{0,0,1,0.1}
\definecolor{Brown}{rgb}{0.396,0.263,0.129}
\definecolor{grey}{cmyk}{0,0,0,0.75}
\definecolor{tangerine}{cmyk}{0,0.5,1,0}
\definecolor{darkgreen}{cmyk}{1,0,1,0.2}

\usepackage[compat=1.0.0]{tikz-feynman}
 \tikzfeynmanset{compat=1.0.0} 
\usepackage{tikz}
\usetikzlibrary{decorations.pathmorphing,decorations.markings}

\usepackage{graphicx}

\tikzset{
photon/.style={decorate, decoration={snake,amplitude=2pt, segment length=5pt}, draw=black},
particle/.style={draw=black, postaction={decorate}, decoration={markings,mark=at position .5 with {\arrow[draw=black]{>}}}},
antiparticle/.style={draw=black, postaction={decorate}, decoration={markings,mark=at position .5 with {\arrowreversed[draw=black]{>}}}},
gluon/.style={decorate, draw=black, decoration={coil,amplitude=4pt, segment length=5pt}},
goldstone/.style={draw=green,postaction={decorate},decoration={markings,mark=at position .5 with {\arrow[draw=blue]{>}}}}
}

\begin{document}

\title{\hfill ~\\[-30mm]
\begin{footnotesize}
\hspace{100mm}
DIAS-STP-24-17\\
\end{footnotesize}
\vspace{3mm}
                  \textbf{On the CP Properties of Spin-0 Dark Matter}        }

\author{\\[-5mm]
A.~Dey\footnote{E-mail: {\tt atri@stp.dias.ie}}$^{~1}$, \
J.~Hern\'andez-S\'anchez\footnote{E-mail: {\tt jaime.hernandez@correo.buap.mx}}$^{~2}$,\ 
V.~Keus\footnote{E-mail: {\tt venus@stp.dias.ie}}$^{~1,3}$,
S.~Moretti\footnote{E-mail: {\tt stefano@soton.ac.uk, stefano.moretti@physics.uu.se}} $^{4,5,6}$,
T.~Shindou\footnote{E-mail: {\tt shindou@cc.kogakuin.ac.jp}} $^{7}$
\\
\emph{\small  $^1$  Dublin Institute for Advanced Studies, School of Theoretical Physics,}\\
\emph{\small 10 Burlington road, Dublin, D04 C932, Ireland}\\
\emph{\small $^2$ Instituto de F\'isica and Facultad de Ciencias de la Electr\'onica,}\\  
\emph{\small 
Benem\'erita Universidad Aut\'onoma de Puebla,
Apdo. Postal 542, C.P. 72570 Puebla, M\'exico,}\\
\emph{\small $^3$ Department of Physics and Helsinki Institute of Physics,}\\
\emph{\small Gustaf Hallstromin katu 2, FIN-00014 University of Helsinki, Finland}\\
\emph{\small $^4$ School of Physics and Astronomy, University of Southampton,}\\
\emph{\small Southampton, SO17 1BJ, United Kingdom}\\
\emph{\small  $^5$ Particle Physics Department, Rutherford Appleton Laboratory,}\\
\emph{\small Chilton, Didcot, Oxon OX11 0QX, United Kingdom}\\
\emph{\small $^6$ Department of Physics and Astronomy, Uppsala University,}\\
\emph{\small Box 516, SE-751 20 Uppsala, Sweden}\\
\emph{\small $^7$ Division of Liberal-Arts, Kogakuin University,}\\
\emph{\small  	2665-1 Nakano-machi, Hachioji, Tokyo, 192-0015, Japan}\\[1mm]
  }

\date{\today}

\maketitle

\begin{abstract}
\noindent
{Aiming to uncover the CP properties of spin-0 particle Dark Matter (DM), we explore a two-component DM scenario within the framework of 3-Higgs Doublet Models (3HDMs), a well-motivated set-up previously studied due to the complementarity of its collider and astrophysical probes. We devise benchmark points in which the two components of DM have same CP  in one case and opposite CP in another. We then show several cross section distributions of observables at collider experiments where the two cases are clearly distinguishable.
} 
 \end{abstract}
\thispagestyle{empty}
\vfill
\newpage
\setcounter{page}{1}

\newpage

\section{Introduction}

The Standard Model (SM) of particle physics has undergone rigorous testing in recent decades,
culminating in 2012 with the discovery of a spin-0 boson by the ATLAS and CMS experiments at the CERN Large Hadron Collider (LHC). This particle, with a mass of approximately 125 GeV, has been identified as the last missing component of the SM \cite{Aad:2012tfa,Chatrchyan:2012ufa}.
Despite the fact that the properties of the observed scalar align with those expected of the SM-Higgs boson, the possibility remains that it could be part of an extended scalar sector.

Although no evidence for physics Beyond the SM (BSM) has yet emerged, it is clear that the SM is not the complete theory of Nature. One significant limitation is its inability to provide a viable candidate for Dark Matter (DM). According to the standard cosmological $\Lambda$CDM Model \cite{Ade:2015xua}, DM should be a particle that is stable over cosmic time-scales, cold (non-relativistic at the onset of galaxy formation), non-baryonic, neutral and weakly interacting, criteria that are not fulfilled by any of the particles contained in the SM.

Several theoretical candidates have been proposed, with Weakly Interacting Massive Particles (WIMPs) \cite{Jungman:1995df, Bertone:2004pz, Bergstrom:2000pn} being among the most studied. These particles are predicted to have masses ranging from a few GeV to a few TeV. Nevertheless, the precise nature of these  DM particle remains elusive.

WIMPs generally exhibit stability due to the preservation of a specific discrete symmetry. For instance, neutralinos, which are prominent candidates for supersymmetric DM, remain stable because of the R-parity conservation \cite{Nilles:1983ge,Haber:1984rc}. In models featuring Universal Extra Dimensions, bosonic DM candidates gain stability through KK-parity, a remnant of momentum conservation in the extra dimension \cite{Cheng:2002ej,Servant:2002aq}. Additionally, in non-minimal Higgs scenarios, scalar DM candidates are stabilised by a conserved $Z_N$ symmetry within the scalar potential \cite{McDonald:1993ex,Burgess:2000yq,Deshpande:1977rw, Ma:2006km,Belanger:2012zr,Barbieri:2006dq,LopezHonorez:2006gr,Ivanov:2012hc}.

A straightforward model that offers a spin-0 DM candidate is the Inert Doublet Model (IDM) \cite{Deshpande:1977rw}. This model includes one inert doublet and one Higgs doublet, also referred to as I(1+1)HDM, and has been the focus of significant research in recent years (see, e.g., \cite{Ma:2006km,Barbieri:2006dq,LopezHonorez:2006gr}). In the IDM, the additional $SU(2)_W$ scalar doublet shares the same quantum numbers as the SM Higgs doublet. One possible vacuum configuration for this model is $(v,0)$, where the first doublet acquires a non-zero Vacuum Expectation Value (VEV) and is termed the \textit{active} doublet. The second doublet, known as the \textit{inert} doublet, does not develop a VEV and thus does not participate in Electro-Weak Symmetry Breaking (EWSB). Since the inert doublet does not couple with fermions and is uniquely $Z_2$-odd in the model, it provides a stable DM candidate: the lightest neutral $Z_2$-odd particle.

The I(1+1)HDM can be viewed as a specific instance of the Higgs-portal DM model, where the interaction between the DM sector and the SM occurs through Higgs boson exchange \cite{Patt:2006fw,Chu:2011be,Queiroz:2014yna}. Consequently, the coupling between DM and the Higgs boson, denoted as $g_{h{\rm DM}{\rm DM}}$, influences the DM annihilation rate $\langle\sigma v\rangle$, the DM-nucleon scattering cross section $\sigma_{{\rm DM}-N}$, and the invisible decays of the Higgs boson. Addressing all these experimental constraints simultaneously can be quite challenging, as demonstrated in studies such as \cite{Mambrini:2011ik,Djouadi:2011aa,Djouadi:2012zc}.

One approach to overcoming this challenge is to break the direct relationship between the annihilation rate and the direct detection cross section by incorporating co-annihilation processes involving DM and other nearly-mass-degenerate inert particles. These co-annihilation processes can alter the effective annihilation cross section, thereby affecting the DM relic density. 
For instance, in the I(1+1)HDM, the DM candidate might co-annihilate with neutral or charged $Z_2$-odd particles. 
Models with a richer particle spectrum could accommodate more co-annihilation processes. An effective way to introduce such additional processes is by including an extra inert singlet \cite{Belanger:2012vp,Yaguna:2019cvp,Belanger:2020hyh} or an extra inert doublet \cite{Ivanov:2012hc, Aranda:2019vda}. In the context of the 3-Higgs Doublet Model (3HDM), also known as the I(2+1)HDM \cite{Keus:2014jha,Keus:2014isa,Keus:2015xya,Cordero-Cid:2016krd,Cordero:2017owj,Cordero-Cid:2018man,Keus:2019szx,Cordero-Cid:2020yba,Keus:2021dti} in the above configuration, such additions can effectively address these constraints.

Proposed by Weinberg in 1976 \cite{Weinberg:1976hu},  3HDMs represent a natural extension beyond the simpler 2HDMs and are highly motivated. All potential finite symmetries in 3HDMs have been thoroughly explored \cite{Ivanov:2012fp}. These models offer a promising framework for addressing the origin and nature of the three fermion families. Specifically, the symmetry associated with the three Higgs doublets might reflect the symmetry governing the three families of quarks and leptons.

In Ref. \cite{Keus:2013hya}, we analysed a symmetric 3HDMs and established the conditions under which the vacuum alignments $(0,0,v_3)$, $(0,v_2,v_3)$, and $(v_1,v_2,v_3)$ minimise the potential. In the present context, we focus on the alignment $(0,0,v_3)$, which is particularly noteworthy due to its similarity to the I(1+1)HDM and the absence of Flavour Changing Neutral Currents (FCNCs).

The I(1+1)HDM continues to be a relevant model for a scalar DM candidate. However, the range of parameter space that aligns with current experimental constraints has significantly diminished \cite{Arina:2009um,Nezri:2009jd,Miao:2010rg,Gustafsson:2012aj,Arhrib:2012ia,Krawczyk:2013pea,Goudelis:2013uca,Arhrib:2013ela,Krawczyk:2015vka,Ilnicka:2015jba,Diaz:2015pyv,Modak:2015uda,Queiroz:2015utg,Garcia-Cely:2015khw,Hashemi:2016wup,Poulose:2016lvz,Alves:2016bib,Datta:2016nfz,Belyaev:2016lok,Belyaev:2018ext,Sokolowska:2019xhe,Kalinowski:2019cxe}. Currently, two viable DM mass regions remain: a low mass region, 53 GeV $\lesssim m_{\text{\rm DM}} \lesssim m_W$, and a heavy mass region $m_{\text{\rm DM}}\lesssim$ 525 GeV. 
In all the region between $m_W$ and $~ 525$ GeV, the DM candidate annihilates very efficiently, giving a total relic density below the observations.
In order to revive this DM region and give a comprehensive model of DM, one needs to invoke the presence of a least a second DM candidate.

In our earlier research \cite{Hernandez-Sanchez:2020aop,Hernandez-Sanchez:2022dnn}, we explored a two-component DM model within the framework of an I(2+1)HDM that is symmetric under a $Z_2 \times Z'_2$ group. In this set-up, one inert doublet is odd under $Z_2$ but even under $Z'_2$, while the other is the reverse. The lightest particle from each inert doublet serves as a viable DM candidate, each possessing a different discrete parity that works together to produce the correct relic density. We demonstrated that additional dark particles from both doublets play a crucial role in determining the final relic abundances of these stable particles by affecting the thermal evolution and decoupling rates of the DM particles. 
A similar study was conducted within a supersymmetric framework in \cite{Khalil:2020syr}.

We showed that, if there is a significant mass difference between the two DM candidates, both can be tested in current and future experiments, as their masses are generally near the EW scale. Specifically, the lighter DM component can be detected through nuclear recoil in direct detection experiments while the heavier DM component can be observed via its contribution to the photon flux in indirect detection experiments.
Additionally, we investigated collider signatures of the I(2+1)HDM, independently of astrophysical probes, by focusing on scalar cascade decays resulting in $\ell^+\ell^- + \Et$ final states ($\ell=e,\mu$ and $\Et$ being missing transverse energy) at the LHC. We analysed several observable distributions, which revealed patterns indicative of the presence of the two distinct DM candidates.

In this paper, we aim to show that when, two spin-0 DM candidates are present, such as in the $Z_2\times Z'_2$-symmetric I(2+1)HDM, one can infer the CP properties of the two DM candidates with respect to each other.
The outline of the paper is as follows. 
In section \ref{sec:model}, we present the scalar potential and the mass spectrum. 
In sections \ref{sec:parameters} and \ref{sec:constraints}, we construct our benchmark scenarios and discuss the available parameter space of the model.  
In section \ref{sec:collider} we present our collider analysis and finally draw our conclusions in section \ref{sec:conclusion}.

\section{The $Z_2 \times Z_2$-Symmetric Model}
\label{sec:model}

The most general $Z_2\times Z'_2$ symmetric 3HDM potential has the following form \cite{Ivanov:2011ae,Keus:2013hya}:  

\bea
\label{potential}
V &=& V_0+V_{Z_2 \times Z'_2},\\[1mm]
V_0 &=& - \mu^2_{1} (\phi_1^\dagger \phi_1) -\mu^2_2 (\phi_2^\dagger \phi_2) - \mu^2_3(\phi_3^\dagger \phi_3) \nonumber
+ \lambda_{11} (\phi_1^\dagger \phi_1)^2+ \lambda_{22} (\phi_2^\dagger \phi_2)^2  + \lambda_{33} (\phi_3^\dagger \phi_3)^2 \nonumber\\
&& + \lambda_{12}  (\phi_1^\dagger \phi_1)(\phi_2^\dagger \phi_2)
 + \lambda_{23}  (\phi_2^\dagger \phi_2)(\phi_3^\dagger \phi_3) + \lambda_{31} (\phi_3^\dagger \phi_3)(\phi_1^\dagger \phi_1) \nonumber\\
&& + \lambda'_{12} (\phi_1^\dagger \phi_2)(\phi_2^\dagger \phi_1) 
 + \lambda'_{23} (\phi_2^\dagger \phi_3)(\phi_3^\dagger \phi_2) + \lambda'_{31} (\phi_3^\dagger \phi_1)(\phi_1^\dagger \phi_3),  \nonumber\\[1mm]
V_{Z_2 \times Z'_2}&=&  \lambda_1 (\phi_1^\dagger \phi_2)^2 + \lambda_2(\phi_2^\dagger \phi_3)^2 + \lambda_3 (\phi_3^\dagger \phi_1)^2 + \mathrm{h.c.}, \nonumber 
\eea
where $V_0$ is invariant under any phase rotation while $V_{Z_2 \times Z'_2}$ ensures the symmetry under the $Z_2 \times Z'_2$ group generated by
\be 
g_{Z_2} = \mathrm{diag}(-1,1,1)\, , \qquad
g_{Z'_2} = \mathrm{diag}(1,-1,1) \,.
\ee
Under this charge assignment, the doublet $\phi_3$ is even under both $Z_2$ and $Z_2'$, while $\phi_1$ is odd under $Z_2$ and even under $Z_2'$ and vice versa for 
$\phi_2$.   
We assign all SM gauge and matter fields an even charge under the $Z_2 \times Z'_2$ symmetry.
In order to preserve our DM candidates the $Z_2 \times Z'_2$ symmetry remains unbroken by  the vacuum. 
In such a case, only $\phi_3$ can develop a VEV and 
we can identify $\phi_3$ with the SM Higgs doublet.

With this setup, 
the Yukawa interactions are set to ``Type-I'' interactions, i.e.,  only the third doublet, $\phi_3$, couples to fermions:
\be
\mathcal{L}_{Y} = \Gamma^u_{mn} \bar{q}_{m,L} \tilde{\phi}_3 u_{n,R} + \Gamma^d_{mn} \bar{q}_{m,L} \phi_3 d_{n,R}
 +  \Gamma^e_{mn} \bar{l}_{m,L} \phi_3 e_{n,R} + \Gamma^{\nu}_{mn} \bar{l}_{m,L} \tilde{\phi}_3 {\nu}_{n,R} + \mathrm{h.c.}  \label{yukawa}
\ee
The $Z_2\times Z_2'$ symmetry forbids the Yukawa 
interactions of the first and second doublets, $\phi_1$ and $\phi_2$, with fermions.
This  ensures that there are indeed no FCNCs.

In a 3HDM with two inert doublets, there are the possible 
effects of dark CP-violation\cite{Cordero-Cid:2016krd,Keus:2016orl,Cordero:2017owj,Cordero-Cid:2018man,Cordero-Cid:2020yba}.
In the model with $Z_2 \times Z_2'$ symmetry, 
all the complex phases in the scalar potential are 
rephased out by the phase redefinition 
of the scalar fields without loss of generality.
Even in the potential without complex phases,
physical CP phase can be induced by the 
spontaneously breaking of the $Z_2 \times Z_2'$ symmetry, 
for particular choices of parameters,
as discussed in \cite{Hernandez-Sanchez:2020aop}.
However, in this paper, 
we consider the case without the spontaneous breaking of the $Z_2 \times Z_2'$ symmetry
so that the CP symmetry is never broken in the scalar sector.

\subsection{Mass Spectrum and Physical Parameters}
\label{sec:mass}

The vacuum alignment $(0,0,v)$ respects the $Z_2\times Z_2'$ symmetry of the potential in which the 
doublets have the following composition:
\be 
\phi_1= \doublet{$\begin{scriptsize}$ H^+_1 $\end{scriptsize}$}{\frac{H_1+iA_1}{\sqrt{2}}} ,\quad 
\phi_2= \doublet{$\begin{scriptsize}$ H^+_2 $\end{scriptsize}$}{\frac{H_2+iA_2}{\sqrt{2}}} , \quad 
\phi_3= \doublet{$\begin{scriptsize}$ H^+_3 $\end{scriptsize}$}{\frac{v+h+iA^0_3}{\sqrt{2}}} \, ,
\label{vac-inert}
\ee
with the extremum condition for this state reading as
\be 
v^2=\frac{\mu_3^2}{\lambda_{33}}\,.
 \label{inertex}
\ee
The third doublet, $\phi_3$, plays the role of the SM-Higgs doublet, with the Higgs particle $h$ having, by construction, tree-level interactions with gauge bosons and fermions identical to those of the SM Higgs boson. Its mass is fixed through the tadpole conditions to be
\be
m^2_h= 2\mu_3^2 = 2 v^2 \lambda_{33} \, =\, (125 \, \mathrm{ GeV})^2 \,,
\ee
and the $A^0_3$ and ${H}^\pm_3$ states are the would-be Goldstone bosons. 

The two inert doublets, $\phi_1$ and $\phi_2$, each contains 
two neutral particles $H_i$ and $A_i$, and one charged particle $H^\pm_i$ with $i=1$ referring to the first doublet and $i=2$ to the second doublet. From here onwards, we refer to particles from the first inert doublet as the first family and to the particles from the second inert doublet as the second family. The masses of the two families are calculated to be
\bea
&& m^2_{H_1}= -\mu^2_1  +\frac{1}{2}(\lambda_{31}+\lambda'_{31} +2\lambda_3 )v^2 \equiv -\mu_1^2 + \Lambda_3 v^2, \label{massmh1}\\
&& m^2_{A_1}= -\mu^2_1  +\frac{1}{2}(\lambda_{31}+\lambda'_{31} -2\lambda_3)v^2 \equiv -\mu_1^2 + \bar{\Lambda}_3 v^2,\\
&& m^2_{H^\pm_1}= -\mu^2_1 +\frac{1}{2}\lambda_{31}v^2
\eea
and
\bea 
&& m^2_{H_2}= -\mu^2_2  +\frac{1}{2}(\lambda_{23}+\lambda'_{23} +2\lambda_2)v^2\equiv -\mu_2^2 + \Lambda_2 v^2, \label{massmh2}\\
&& m^2_{A_2}= -\mu^2_2  +\frac{1}{2}(\lambda_{23}+\lambda'_{23} -2\lambda_2)v^2\equiv -\mu_2^2 + \bar{\Lambda}_2 v^2,\\
&& m^2_{H^\pm_2}= -\mu^2_2 +\frac{1}{2}\lambda_{23} v^2. \label{massmhc2}
\eea

For our analysis, we rewrite the parameters of the potential in terms of  physical observables, such as masses and couplings. The tree-level SM couplings in the gauge and fermionic sectors follow exactly the SM definitions. The relevant parameters arising from the extended scalar sector are: (i) masses of inert particles and the Higgs-DM couplings, which represent parameters from the visible sector; (ii) self-interaction parameters, which describe interactions within the dark sector. The full list is:
\be
v^2, \, m_h^2, \, m^2_{H_1},\,  m^2_{H_2}, \, m^2_{A_1}, \, m^2_{A_2}, \, m^2_{H^\pm_1}, \, m^2_{H^\pm_2}, \, \Lambda_2, \, \Lambda_3,\,  \Lambda_1, \, \lambda_{11},\,  \lambda_{22},\,  \lambda'_{12}, \, \lambda_{12} \label{physpar}.
\ee
The self-couplings $\lambda_{11}, \lambda_{22}, \lambda'_{12}, \lambda_{12}$ correspond exactly to the terms in Eq.  (\ref{potential}), while the relations between remaining parameters and our chosen physical basis are as follows.
\bea
\mu_1^2 &=& -m_{H_1}^2+\Lambda_{3}v^2, \\[1mm]
\lambda_{3} &=& (m_{H_1}^2-m_{A_1}^2)/(2v^2), \\[1mm]
\lambda_{31}' &=& (m_{H_1}^2+m_{A_1}^2-2m_{H^\pm_1}^2)/v^2, \\[1mm]
\lambda_{31} &=& 2\Lambda_{3}-2\lambda_{3}-\lambda_{31}',\\[1mm]
\mu_2^2 &=& -m_{H_2}^2+\Lambda_{2}v^2, \\[1mm]
\lambda_{2} &=& (m_{H_2}^2-m_{A_2}^2)/(2v^2),\\[1mm]
\lambda_{23}' &=& (m_{H_2}^2+m_{A_2}^2-2m_{H^\pm_2}^2)/v^2, \\[1mm]
\lambda_{23} &=& 2\Lambda_{2}-2\lambda_{2}-\lambda_{23}',\\[1mm]
\lambda_1  &=&  2\Lambda_1 - (\lambda_{12}+\lambda'_{12}).
\eea

In principle, any particle among $(H_i, A_i, H^\pm_i)$ can be the lightest. Here, we dismiss the possibility of $H^\pm_i$ being the lightest, as it would mean that the DM candidate is a charged particle. Choosing between $H_1$ and $A_1$ (or $H_2$ and $A_2$) is related only to a change of the sign of the quartic parameter $\lambda_{3}$ ($\lambda_2$) and has no impact on the ensuing phenomenology. 
As will be discussed in detail in the next section, without loss of generality, we choose $H_1$ to be the DM candidate in the first family, therefore
\be 
m_{H_1} < m_{A_1},\, m_{H_1^\pm} \, \quad \Longrightarrow  \quad \lambda_3<0, \;\; \lambda_{31}'+ 2\lambda_3<0,
\label{eq:1st-family-H1}
\ee
Within the second family, for $H_2$ to be the DM candidate, one requires
\be 
m_{H_2} < m_{A_2},\, m_{H_2^\pm} \, \quad \Longrightarrow \quad \lambda_2<0, \;\; \lambda_{23}'+ 2\lambda_2<0,
\label{eq:2nd-family-H2}
\ee
and for $A_2$ to be the DM candidate, one requires
\be 
m_{A_2} < m_{H_2},\, m_{H_2^\pm} \, \quad \Longrightarrow \quad \lambda_2>0, \;\; \lambda_{23}'- 2\lambda_2<0,
\label{eq:2nd-family-A2}
\ee
Notice that, unlike many $Z_N$ symmetric models, the two lightest states from two doublets are automatically stable, regardless of their mass hierarchy, as they are stabilised by different $Z_2$ symmetries.

\section{Construction of our Benchmark Scenarios}
\label{sec:parameters}

A detailed DM analysis of the $Z_2 \times Z'_2$ symmetric I(2+1)HDM was presented in \cite{Hernandez-Sanchez:2020aop,Hernandez-Sanchez:2022dnn}, with an emphasis on the astrophysical and collider probes of the model and the complementarity between these.
For all our analyses, we use the \texttt{micrOMEGAs} package~\cite{Belanger:2006is} to calculate the relic density of the two DM candidates. For our analysis, we have used the standard assumptions included in the code, {i.e.}, the particles within the dark sector are in thermal equilibrium and have the same kinetic temperature as the SM particle bath and that the DM particles \textit{freeze-out} when their number density multiplied by their annihilation cross section becomes too small compared to the expansion rate of the Universe.  

In principle, the lightest particle in each family is a viable DM candidate. As mentioned before, in the construction of our benchmark scenarios, we take $H_1$ from the first family to be one of the DM candidates and either $H_2$ or $A_2$ from the second family to be the other DM component. 
Note that, in any case, other dark particles from both families have a significant impact on the final relic abundances of these two DM components, as they influence the thermal evolution and decoupling rate of DM particles.
In particular, there are two distinct classes of processes that can influence number densities of dark particles in each sector, namely, (co)annihilations and conversions\footnote{Due to the imposed symmetry there are no processes that would be classified as semi-annihilation, {i.e.}, processes of the form $S_i  S_j \to \textrm{SM } S_k$.}.

\paragraph{(Co)annihilations}
This class contains processes of the type 
$S_i \, S_i \to \textrm{ SM SM}$, 
where $Si$ represents any of the dark particles $H_i$, $A_i$, or $H_i^\pm$ from each family. Specifically, we focus on the standard DM annihilation processes $H_i \, H_i \rightarrow \text{SM} \, \text{SM}$ (for $i = 1, 2$), where the outcomes are heavily influenced by the DM particle masses. For DM particles with masses $m_{\rm DM} \lesssim m_h / 2$, annihilation predominantly proceeds through Higgs-mediated channels into fermions. In contrast, for heavier DM particles, annihilation primarily occurs into gauge boson pairs, either directly or via the Higgs $s$-channel. Our analysis also includes contributions from annihilation into virtual gauge bosons, which play a significant role in determining DM annihilation rates for intermediate mass ranges ($m_h / 2 \lesssim m_{\rm DM} \lesssim m_{W^\pm}$). 

Additionally, when the mass difference between the DM candidate and other neutral or charged inert scalars in the same family is small (within 10\% of the DM mass), co-annihilation processes such as $H_i \, A_i \rightarrow Z \rightarrow \text{SM} \, \text{SM}$ become crucial. These co-annihilation channels are particularly dominant for high DM masses ($m_{\rm DM} \gtrsim 500 \text{ GeV}$), similar to what is observed in the I(1+1)HDM.
It is important to note that, due to the $Z_2 \times Z'_2$ imposed symmetry here, the two dark families are separated. As a result, there are no vertices that involve fields from two separate families, namely, Higgs or gauge bosons couple only to a pair of dark particles from the same family. 

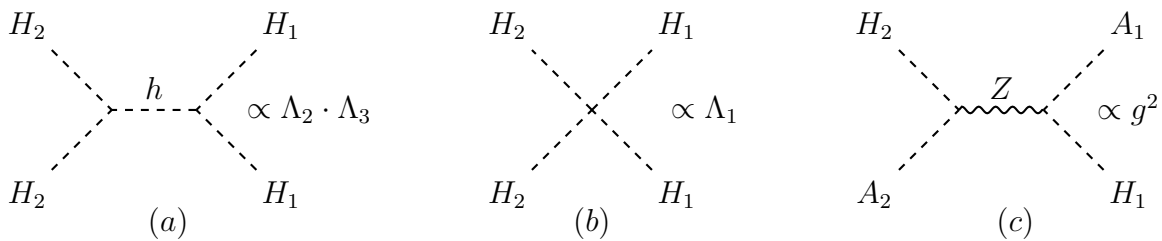
\begin{figure}[h!]
\begin{center}
\begin{tikzpicture}[thick,scale=0.75]
    \begin{feynman}
        \vertex (m) at ( 0, 0);
        \vertex (n) at (1.5,0);
        \vertex (a) at (-1.5,-1.5) {\(H_2\)};
        \vertex (b) at ( 3,-1.5) {\(H_1\)};
        \vertex (c) at (-1.5, 1.5) {\(H_2\)};
        \vertex (d) at ( 3, 1.5) {\(H_1\)};
        \vertex (e) at ( 1, -2) {\((a)\)};  
		\vertex (f) at ( 3.5, 0) {\(\propto \Lambda_2 \cdot \Lambda_3\)};  
        \diagram* {
            (a) -- [scalar] (m),
            (b) -- [scalar] (n),
            (c) -- [scalar] (m),
            (d) -- [scalar] (n),
			(n) -- [scalar, edge label'=\(h\)] (m),
        };
    \end{feynman}
\end{tikzpicture} 
\hspace{10mm}
\begin{tikzpicture}[thick,scale=0.75]
    \begin{feynman}
        \vertex (m) at ( 0, 0);
        \vertex (a) at (-1.5,-1.5) {\(H_2\)};
        \vertex (b) at ( 1.5,-1.5) {\(H_1\)};
        \vertex (c) at (-1.5, 1.5) {\(H_2\)};
        \vertex (d) at ( 1.5, 1.5) {\(H_1\)};
        \vertex (e) at ( 0, -2) {\((b)\)};        
		\vertex (f) at ( 2, 0) {\(\propto \Lambda_1\)};  
        \diagram* {
            (a) -- [scalar] (m) -- [scalar] (c),
            (b) -- [scalar] (m) -- [scalar] (d),
        };
    \end{feynman}
\end{tikzpicture}
\hspace{10mm}
\begin{tikzpicture}[thick,scale=0.75]
    \begin{feynman}
        \vertex (m) at ( 0, 0);
        \vertex (n) at (1.5,0);
        \vertex (a) at (-1.5,-1.5) {\(A_2\)};
        \vertex (b) at ( 3,-1.5) {\(H_1\)};
        \vertex (c) at (-1.5, 1.5) {\(H_2\)};
        \vertex (d) at ( 3, 1.5) {\(A_1\)};
        \vertex (e) at ( 1, -2) {\((c)\)};  
		\vertex (f) at ( 3, 0) {\(\propto g^2\)};  
        \diagram* {
            (a) -- [scalar] (m),
            (b) -- [scalar] (n),
            (c) -- [scalar] (m),
            (d) -- [scalar] (n),
			(n) -- [boson, edge label'=\(Z\)] (m),           };
    \end{feynman}
\end{tikzpicture} 
\caption{Examples of DM conversion diagrams include: (a) Higgs-mediated conversion of $H_2 H_2 \to H_1 H_1$, which is always present as long as $\Lambda_{2,3} \neq 0$; (b) direct conversion between DM particles, which depends on the self-interaction parameter $\Lambda_1$; and (c) $Z$-mediated conversion resulting from co-annihilation processes, which depends on the gauge coupling $g$.}
 \label{fig-conversion}
 \end{center}
\end{figure}

\paragraph{Conversions}
In this class of processes, a pair of heavier dark particles from one family can convert into a pair of dark particles from another family, $S_i \, S_i \rightarrow S_j \, S_j$, either directly or via interaction with an SM particle. It is important to note that such conversions between different families of dark particles can occur even when all self-interaction couplings are set to zero. In Fig.~\ref{fig-conversion}, diagrams (a) and (b) illustrate the conversion of a pair of $H_2$ particles into a pair of $H_1$ particles, either through Higgs-mediated processes or direct conversion. 

Even if the self-interaction parameter $\Lambda_1$ were zero, diagram (a) would still contribute to the conversion process, provided that both particles interact with the Higgs boson ($\Lambda_{2,3} \neq 0$). We anticipate that the overall contribution could be subject to cancellations or enhancements depending on the relative signs of $\Lambda_2 \Lambda_3$ and $\Lambda_1$. Additionally, depending on the masses and coupling constants, we also consider processes where heavier dark particles from the second family annihilate directly into stable particles from the first family, such as $A_2 A_2 \rightarrow h \rightarrow H_1 H_1$. All these processes are automatically incorporated into our numerical analysis.

As previously mentioned, the parameters of the potential in Eq.~(\ref{physpar}) can be expressed in terms of the masses of the scalar particles and their couplings. The masses of the inert/dark particles, $m^2_{H_1}, m^2_{H_2}, m^2_{A_1}, m^2_{A_2}, m^2_{H^\pm_1}, m^2_{H^\pm_2}$, determine the annihilation behaviour of the DM particles. Based on the absolute values of these masses as well as the mass splittings between particles, different dominant channels for annihilation, co-annihilation and conversion can be expected.
Additionally, the absolute and relative values of these masses will give rise to different potential collider signatures.

The self-couplings of the dark sector particles, denoted by $\lambda_1, \lambda'_{12}, \lambda_{12}$, $\lambda_{11}$ and $ \lambda_{22}$, serve two distinct purposes. The first set governs interactions between particles from different families and significantly influences the DM relic density via the conversion processes. Specifically, the following couplings play a key role in shaping DM phenomenology:
\bea
g_{H_1H_1H_2H_2} &=&  \phantom{-} 2\lambda_1 +\lambda_{12} + \lambda'_{12} =  4\Lambda_1 -(\lambda_{12} + \lambda'_{12}) \,, \\
g_{A_1A_1H_2H_2}  = g_{A_2A_2H_1H_1} &=&  -4\lambda_1 +\lambda_{12} + \lambda'_{12} = -4\Lambda_1 + 2(\lambda_{12} + \lambda'_{12}) \,.
\eea
In contrast, the couplings $\lambda_{11}$ and $\lambda_{22}$ do not directly impact any observable processes, nor do they affect the DM relic density or collider signals. Nevertheless, they are essential in determining the allowed parameter space due to their role in ensuring vacuum stability.

The couplings between the DM particles and the Higgs boson, $\Lambda_2$ and $\Lambda_3$, play a crucial role in governing not only DM annihilation and conversion processes but also affect the possible invisible decays of the Higgs boson as well as direct and indirect DM detection. In our numerical study, we identify  the following interaction vertices, in particular, as having a notable impact on the overall DM phenomenology:
\bea
g_{hH_1H_1} &=& \phantom{-} 2\lambda_3 +\lambda_{31} + \lambda'_{31} = 2 \Lambda_3 \,, 
\\[1mm]
g_{hA_1A_1} &=& -2\lambda_3 +\lambda_{31} + \lambda'_{31} = 2\Lambda_3 + 2(m^2_{A_1} - m^2_{H_1})/v^2  \,,  
\label{Eq:cubic-couplings}
\\[1mm]
g_{hH_2H_2} &=& \phantom{-} 2\lambda_2 +\lambda_{23} + \lambda'_{23} = 2 \Lambda_2 \,, \\[1mm]
g_{hA_2A_2} &=&  -2\lambda_2 +\lambda_{23} + \lambda'_{23} = 2\Lambda_2 + 2(m^2_{A_2} - m^2_{H_2})/v^2 \,.
\eea

Let  us point out that, for a given mass splitting of the neutral fields within a family, the coupling of Higgs to a pair of neutral particles could significantly change depending on the mass ordering:
\be 
g_{hA_iA_i}  - g_{hH_iH_i}   = \left\lbrace  \begin{array}{l} 
+ \Delta_i \quad \mathrm{if} \,\, m_{H_i}<m_{A_i}\\[2mm]
- \Delta_i  \quad  \mathrm{if} \,\, m_{A_i}<m_{H_i}
\end{array}\right.\\
\qquad \mathrm{where} \quad  \Delta_i= 2 \left| m^2_{A_i} - m^2_{H_i}    \right| /v^2   \,.
\ee
To clarify the importance of this statement, consider the following set-up.
Within the first family, we take the CP-even state $H_1$ to be the DM candidate, therefore, 
\be 
g_{hA_1A_1}  - g_{hH_1H_1}= +\Delta_1 \quad \Rightarrow \quad g_{hA_1A_1}  > g_{hH_1H_1}\,.
\ee
This means that, in the regions of the parameter space where the $H_1$-$A_1$ mass splitting is large\footnote{Which is satisfied when $|\lambda_3|$ is large while in agreement with all theoretical and experimental bounds.} (meaning $\Delta_1$ is large), the co-annihilation processes between $H_1$ and $A_1$ are not very efficient. As a result, the annihilation cross section, proportional to $g_{hH_1H_1}=2\Lambda_3$, must be sufficiently large to avoid over-closing the universe. 
To conclude, when the CP-even (scalar) state  is the DM candidate, if the Higgs coupling to the CP-even particle, $g_{hH_1H_1}$, is positive, then the Higgs coupling to the CP-odd particle $g_{hA_1A_1}$ will also be positive. 

Now, consider a case where the CP-odd particle $A_1$ is the DM candidate, hence, 
\be 
g_{hA_1A_1}  - g_{hH_1H_1}= -\Delta_1 \quad \Rightarrow \quad g_{hA_1A_1}  < g_{hH_1H_1}\,,
\ee
which means that, in the regions of the parameter space where the $A_1$-$H_1$ mass splitting is large and as a result, the $H_1$-$A_1$ co-annihilation processes are not efficient, the annihilations of $A_1$ through Higgs, proportional to $g_{hA_1A_1} = 2\Lambda_3 - \Delta_1$ should be large\footnote{Conversion processes could also contribute to reducing the relic abundance of $A_1$ provided the particles from the second family are lighter than $A_1$. These processes are again proportional to $g_{hA_1A_1}$ and, therefore, similar to the annihilation processes, benefit from a large $g_{hA_1A_1}$. A large $\Lambda_1$ could also enhance the conversion processes through the direct conversion diagram. If the particles of the second family are not lighter than the particles in the first family, the conversion process will have a sub-dominant effect.}. 
Note that for a sufficiently large $A_1$-$H_1$ mass splitting, {i.e.}, sufficiently large $\Delta_1$, the Higgs coupling to the CP-odd particle $g_{hA_1A_1}$ could be negative, even if $g_{hH_1H_1}$ is positive. We find this to be the case in a large region of the parameter space. 

In constructing our benchmark scenarios, we take advantage of the fact that the sign of the Higgs-DM coupling depends on the CP properties of the DM particles. 
Therefore, we  expect that interference effects between the diagrams carrying these couplings and others would produce contributions to the cross section that   are distinctive of the relative CP status of the two DM candidates in our   chosen BSM scenario.

\subsection{Two Distinct Benchmark Scenarios}\label{sec:bBPs}

We construct our benchmark scenarios with the primary objective to identify distinct signatures that can differentiate between CP properties of the two components of DM, in particular, at collider experiments. 
To this end, we examine two distinct scenarios as follows:
\begin{itemize}
\item {\bf Scenario 1} in which the DM candidates from the two families have  same CP, namely, $H_1$ from the first family and $H_2$ from the second family.
\item {\bf Scenario 2} in which the DM candidates from the two families have  opposite CP, namely, $H_1$ from the first family and $A_2$ from the second family.
\end{itemize} 
We identify a viable Benchmark Point (BP) in each of the two scenarios. Note that our goal is to isolate the effect of the CP of the two DM components with respect to each other. In order to do that, we choose BPs in which the CP states of the  DM components are either the same (in BP1) or the opposite (in BP2) while all other characteristics of the DM components are the same, in particular, as follows.
\begin{itemize}
\item 
In each BP, the two DM components have the same mass: 
\be 
m_{H_1}^{\bts \mathrm{BP1} \ets} = m_{H_2}^{\bts \mathrm{BP1} \ets}\;, \qquad 
m_{H_1}^{\bts \mathrm{BP2} \ets} = m_{A_2}^{\bts \mathrm{BP2} \ets}.
\ee
\item 
The DM candidate from the first(second) family in BP1 has the same mass as the DM candidate from the first(second) family in BP2:
\be 
m_{H_1}^{\bts \mathrm{BP1} \ets} = m_{H_1}^{\bts \mathrm{BP2} \ets} \;, \qquad 
m_{H_2}^{\bts \mathrm{BP1} \ets} = m_{A_2}^{\bts \mathrm{BP2} \ets}.
\ee
\item 
The masses of additional dark scalars in both BPs are the same:
\be 
m_{A_1}^{\bts \mathrm{BP1} \ets} = m_{A_1}^{\bts \mathrm{BP2} \ets} \;, \qquad 
m_{H^\pm_1}^{\bts \mathrm{BP1} \ets} = m_{H^\pm_1}^{\bts \mathrm{BP2} \ets}\;, \qquad 
m_{A_2}^{\bts \mathrm{BP1} \ets} = m_{H_2}^{\bts \mathrm{BP2} \ets} \;, \qquad 
m_{H^\pm_2}^{\bts \mathrm{BP1} \ets} = m_{H^\pm_2}^{\bts \mathrm{BP2} \ets}.
\ee 
\item 
In order for both BPs to lead to similar DM relic density and prodcution/annihilation cross sections, the relevant couplings in both BPs are the same:
\be 
g_{hH_1H_1}^{\bts \mathrm{BP1} \ets} = g_{hH_1H_1}^{\bts \mathrm{BP2} \ets} \;, \qquad 
g_{hH_2H_2}^{\bts \mathrm{BP1} \ets} \simeq |g_{hA_2A_2}^{\bts \mathrm{BP2} \ets} |.
\ee 
\end{itemize}

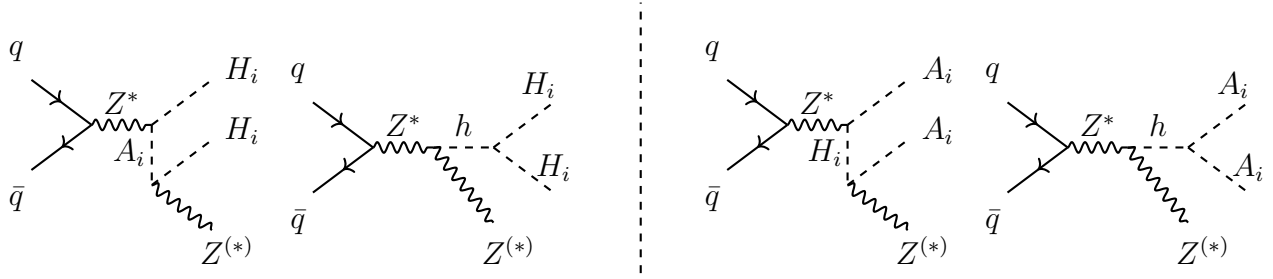
\begin{figure}[t!]
\centering 
\begin{tikzpicture}[thick,scale=0.8]
\draw[particle] (10,0) -- node[black,above,xshift=-0.6cm,yshift=0.4cm] {$q$} (11,-0.75);
\draw[antiparticle] (10,-1.5) -- node[black,above,yshift=-1.0cm,xshift=-0.6cm] {$\bar{q}$} (11,-0.75);
\draw[photon] (11,-0.75) -- node[black,above,xshift=0.0cm,yshift=0.0cm] {$Z^*$} (12,-0.75);
\draw[dashed] (12,-0.75) -- node[black,above,yshift=0.1cm,xshift=0.8cm] {$H_i$} (13,-0);
\draw[dashed] (12,-0.75) -- node[black,above,yshift=-0.3cm,xshift=-0.3cm] {$A_{i}$} (12,-1.75);
\draw[dashed] (12,-1.75) -- node[black,above,yshift=0.1cm,xshift=0.8cm] {$H_i$} (13,-1);
\draw[photon] (12,-1.75) -- node[black,above,xshift=0.6cm,yshift=-0.9cm] {$Z^{(*)}$} (13,-2.5);
\end{tikzpicture}
\begin{tikzpicture}[thick,scale=0.8]
\draw[particle] (10,0) -- node[black,above,xshift=-0.6cm,yshift=0.4cm] {$q$} (11,-0.75);
\draw[antiparticle] (10,-1.5) -- node[black,above,yshift=-1.0cm,xshift=-0.6cm] {$\bar{q}$} (11,-0.75);
\draw[photon] (11,-0.75) -- node[black,above,xshift=0.0cm,yshift=0.0cm] {$Z^*$} (12,-0.75);
\draw[dashed] (12,-0.75) -- node[black,above,yshift=0cm,xshift=0cm] {$h$} (13,-0.75);
\draw[dashed] (13,-0.75) -- node[black,above,yshift=0.2cm,xshift=0.2cm] {$H_i$} (14,0.0);
\draw[dashed] (13,-0.75) -- node[black,above,yshift=-0.3cm,xshift=0.4cm] {$H_i$} (14,-1.5);
\draw[photon] (12,-0.75) -- node[black,above,xshift=0.6cm,yshift=-1.2cm] {$Z^{(*)}$} (13,-2);
\end{tikzpicture}
\hspace{5mm}
\begin{tikzpicture}[thick,scale=0.8]
\draw[dashed] (0,-3) -- node[black,above,yshift=0.2cm,xshift=0.2cm] {} (0,1.5);
\end{tikzpicture}
\hspace{1mm}
\begin{tikzpicture}[thick,scale=0.8]
\draw[particle] (10,0) -- node[black,above,xshift=-0.6cm,yshift=0.4cm] {$q$} (11,-0.75);
\draw[antiparticle] (10,-1.5) -- node[black,above,yshift=-1.0cm,xshift=-0.6cm] {$\bar{q}$} (11,-0.75);
\draw[photon] (11,-0.75) -- node[black,above,xshift=0.0cm,yshift=0.0cm] {$Z^*$} (12,-0.75);
\draw[dashed] (12,-0.75) -- node[black,above,yshift=0.1cm,xshift=0.8cm] {$A_i$} (13,-0);
\draw[dashed] (12,-0.75) -- node[black,above,yshift=-0.3cm,xshift=-0.3cm] {$H_{i}$} (12,-1.75);
\draw[dashed] (12,-1.75) -- node[black,above,yshift=0.1cm,xshift=0.8cm] {$A_i$} (13,-1);
\draw[photon] (12,-1.75) -- node[black,above,xshift=0.6cm,yshift=-0.9cm] {$Z^{(*)}$} (13,-2.5);
\end{tikzpicture}
\begin{tikzpicture}[thick,scale=0.8]
\draw[particle] (10,0) -- node[black,above,xshift=-0.6cm,yshift=0.4cm] {$q$} (11,-0.75);
\draw[antiparticle] (10,-1.5) -- node[black,above,yshift=-1.0cm,xshift=-0.6cm] {$\bar{q}$} (11,-0.75);
\draw[photon] (11,-0.75) -- node[black,above,xshift=0.0cm,yshift=0.0cm] {$Z^*$} (12,-0.75);
\draw[dashed] (12,-0.75) -- node[black,above,yshift=0cm,xshift=0cm] {$h$} (13,-0.75);
\draw[dashed] (13,-0.75) -- node[black,above,yshift=0.2cm,xshift=0.2cm] {$A_i$} (14,0.0);
\draw[dashed] (13,-0.75) -- node[black,above,yshift=-0.3cm,xshift=0.4cm] {$A_i$} (14,-1.5);
\draw[photon] (12,-0.75) -- node[black,above,xshift=0.6cm,yshift=-1.2cm] {$Z^{(*)}$} (13,-2);
\end{tikzpicture}
\caption{The sum of the two diagrams on the left are proportional to $g^2_{ZH_iA_i}$ and $g_{hZZ}  \times g_{hH_iH_i}$, respectively, while the sum of the two diagrams on the right  are proportional to $g^2_{ZH_iA_i}$ and $g_{hZZ}  \times g_{hA_iA_i}$, respectively. With the $g_{hH_iH_i}$ coupling having the opposite sign with respect to the $g_{hA_iA_i}$ one, there will be constructive versus destructive interference between the diagrams contributing to the signal $q \bar{q} \to \ell^+ \ell^- + \Et$. }
\label{fig:Et2l-diags}
\end{figure}


\begin{table} [h!]
\begin{center}
\begin{footnotesize}
\begin{tabular}{|p{5mm}||p{5mm}|p{5mm}|p{6mm}||p{5mm}|p{5mm}|p{6mm}||c|c|c|c|c|c|c|} \hline\\[-2.7ex]
BP & 
$m_{H_1}$ &
$m_{A_1}$ &
$m_{H_1^\pm}$ &
$m_{H_2}$ &
$m_{A_2}$ &
$m_{H_2^\pm}$ &
$\Lambda_1$ &
$g_{hH_1H_1}$ &
$g_{hH_2H_2}$ &
$g_{hA_1A_1}$ &
$g_{hA_2A_2}$ &
$\Omega_{H_1}h^2$ &
$\Omega_{H_2}h^2$
 \\
\hline
BP1 & 
80 &
120 &
130 &
80 &
110 &
130 &
0.082 &
0.192 &
0.18 &
0.2492 &
0.22 &
0.003 &
0.003\\
\hline
\end{tabular}
\\[5mm]
\begin{tabular}{|p{5mm}||p{5mm}|p{5mm}|p{6mm}||p{5mm}|p{5mm}|p{6mm}||c|c|c|c|c|c|c|} \hline\\[-2.7ex]
BP & 
$m_{H_1}$ &
$m_{A_1}$ &
$m_{H_1^\pm}$ &
$m_{H_2}$ &
$m_{A_2}$ &
$m_{H_2^\pm}$ &
$\Lambda_1$ &
$g_{hH_1H_1}$ &
$g_{hH_2H_2}$ &
$g_{hA_1A_1}$ &
$g_{hA_2A_2}$ &
$\Omega_{H_1}h^2$ &
$\Omega_{A_2}h^2$
 \\
\hline
BP2 & 
80 &
120 &
130 &
110 &
80 &
130 &
0.034 & 
0.192 &
0.01 &
0.46 &
-0.18 &
0.004 &
0.005\\
\hline
\end{tabular}
\end{footnotesize}
\caption{\footnotesize The parameter values for BP1 and BP2. In both cases, we have set $\lambda_{11}=0.11$, $\lambda_{22}=0.12$, $\lambda_{12}=0.121$, $\lambda'_{12}=0.13$, the SM Higgs mass $m_{h}=125$ GeV and $v=246$ GeV, are in agreement with all astrophysical and collider constraints. For  BP1, the cross section is $\sigma(e^+ e^- \to \ell^+\ell^- + H_1H_1 / H_2H_2) = 5.9$ fb and, for  BP2, it is $\sigma(e^+ e^- \to  \ell^+\ell^- + H_1H_1 / A_2A_2) = 6.1$ fb for 500 GeV centre-of-mass energy. For BP1, the cross section is $\sigma(e^+ e^- \to  \ell^+\ell^- +  H_1H_1 / H_2H_2) = 2.1$ fb and for the BP2, it is $\sigma(e^+ e^- \to  \ell^+\ell^- + H_1H_1 / A_2A_2) = 1.7$ fb for 1 TeV centre-of-mass energy.}
\label{tab:BPs}
\end{center}
\end{table}

In 
Tab.~\ref{tab:BPs}, we show the details of the two BPs. This configuration allows for a consistent comparison between the scenarios, highlighting the role of coupling signs in differentiating the CP nature of the DM components. 
We aim to determine whether there are any significant differences between these two scenarios that could be observed in future collider experiments, such as an  $e^+ e^-$ linear collider. We focus on the  channel $e^+ e^- \rightarrow \ell^+ \, \ell^- + \mathrm{ DM}\, \mathrm{\rm DM} $, which will give us two opposite sign leptons and missing (transverse) energy in the final state. Note that in BP1, the two DM components are $H_1$ and $H_2$, while in BP2, they are $H_1$ and $A_2$.

The cross sections of the events for the signals are computed at leading order and created using \texttt{Madgraph@MCNLO}~\cite{Alwall:2014hca}. The detector simulation is handled by \texttt{Delphes-3.5.0}~\cite{deFavereau:2013fsa}, wherein we have used the inbuilt detector efficiencies to identify final state isolated particles. We use no further trigger efficiencies and use \texttt{PYTHIA8}~\cite{Sjostrand:2006za} for parton  showering and hadronisation.

\begin{figure}[t!]
\centering 
\begin{tikzpicture}[thick,scale=1.0]
\draw[particle] (10,0) -- node[black,above,xshift=-0.5cm,yshift=0.4cm] {$e^-$} (11,-0.75);
\draw[antiparticle] (10,-1.5) -- node[black,above,yshift=-1.0cm,xshift=-0.6cm] {$e^+$} (11,-0.75);
\draw[photon] (11,-0.75) -- node[black,above,xshift=0.0cm,yshift=0.0cm] {$Z$} (12,-0.75);
\draw[dashed] (12,-0.75) -- node[black,above,yshift=0.1cm,xshift=0.8cm] {$H_i$} (13,-0);
\draw[dashed] (12,-0.75) -- node[black,above,yshift=-0.3cm,xshift=-0.3cm] {$A_i$} (12,-1.75);
\draw[dashed] (12,-1.75) -- node[black,above,yshift=0.1cm,xshift=0.8cm] {$H_i$} (13,-1);
\draw[photon] (12,-1.75) -- node[black,above,xshift=-0.2cm,yshift=-0.5cm] {$Z$} (13,-2.5);
\draw[particle] (13,-2.5) -- node[black,above,xshift=-0.2cm,yshift=0cm] {$e^-$} (14,-1.8);
\draw[antiparticle] (13,-2.5) -- node[black,above,xshift=-0.2cm,yshift=-0.6cm] {$e^+$} (14,-3);
\end{tikzpicture}
\hspace{0.5cm}
\begin{tikzpicture}[thick,scale=1.0]
\draw[particle] (10,0) -- node[black,above,xshift=-0.5cm,yshift=0.4cm] {$e^-$} (11,-0.75);
\draw[antiparticle] (10,-1.5) -- node[black,above,yshift=-1.0cm,xshift=-0.6cm] {$e^+$} (11,-0.75);
\draw[photon] (11,-0.75) -- node[black,above,xshift=0.0cm,yshift=0.0cm] {$Z$} (12,-0.75);
\draw[dashed] (12,-0.75) -- node[black,above,yshift=0cm,xshift=-0.1cm] {$h$} (13,0);
\draw[photon] (12,-0.75) -- node[black,above,yshift=-0.6cm,xshift=-0.1cm] {$Z$} (13,-1.5);
\draw[dashed] (13,0) -- node[black,above,yshift=-0.1cm,xshift=0.8cm] {$H_i$} (14,0.5);
\draw[dashed] (13,0) -- node[black,above,xshift=0.8cm,yshift=-0.5cm] {$H_i$} (14,-0.5);
\draw[antiparticle] (13,-1.5) -- node[black,above,xshift=-0.2cm,yshift=-0.1cm] {$e^+$} (14.1,-1);
\draw[particle] (13,-1.5) -- node[black,above,xshift=-0.2cm,yshift=-0.6cm] {$e^-$} (14.1,-2);
\end{tikzpicture}
\hspace{0.5cm}
\begin{tikzpicture}[thick,scale=1.0]
\draw[particle] (10,0) -- node[black,above,xshift=-0.5cm,yshift=0.4cm] {$e^-$} (11,-0.75);
\draw[antiparticle] (10,-1.5) -- node[black,above,yshift=-1.0cm,xshift=-0.6cm] {$e^+$} (11,-0.75);
\draw[photon] (11,-0.75) -- node[black,above,xshift=0.0cm,yshift=0.0cm] {$Z$} (12,-0.75);
\draw[dashed] (12,-0.75) -- node[black,above,yshift=0.1cm,xshift=1.1cm] {$H_i$} (13.3,0.1);
\draw[dashed] (12,-0.75) -- node[black,above,yshift=-0.2cm,xshift=1.1cm] {$H_i$} (13.5,-0.5);
\draw[photon] (12,-0.75) -- node[black,above,yshift=-0.6cm,xshift=-0.1cm] {$Z$} (13,-1.5);
\draw[antiparticle] (13,-1.5) -- node[black,above,xshift=-0.2cm,yshift=-0.1cm] {$e^+$} (14.1,-1);
\draw[particle] (13,-1.5) -- node[black,above,xshift=-0.2cm,yshift=-0.6cm] {$e^-$} (14.1,-2);
\end{tikzpicture}
\caption{The $s$-channel diagrams leading to the $2\ell + \Et$ final state mediated by the $Z,\, h$ and $A_i$ propagators.}
\label{fig:Et2l-ILC-s}
\end{figure}

\begin{figure}[t!]
\centering 
\begin{tikzpicture}[thick,scale=0.8]
\draw[particle] (10,0.2) -- node[black,above,xshift=-0.9cm,yshift=0.1cm] {$e^-$} (12,0);
\draw[antiparticle] (10,-3.2) -- node[black,above,yshift=-0.7cm,xshift=-0.9cm] {$e^+$} (12,-3.0);
\draw[particle] (12,0) -- node[black,above,xshift=1.1cm,yshift=0.1cm] {$e^-$} (14,0.2);
\draw[antiparticle] (12,-3.0) -- node[black,above,yshift=-0.7cm,xshift=1.1cm] {$e^+$} (14,-3.2);
\draw[photon] (12,0) -- node[black,above,yshift=-0.2cm,xshift=-0.4cm] {$Z$} (12,-1.0);
\draw[dashed] (12,-1) -- node[black,above,yshift=-0.3cm,xshift=-0.4cm] {$A_i$} (12,-2.0);
\draw[photon] (12,-2) -- node[black,above,yshift=-0.4cm,xshift=-0.4cm] {$Z$} (12,-3.0);
\draw[dashed] (12,-1) -- node[black,above,yshift=-0.1cm,xshift=-0.2cm] {$H_i$} (14,-0.8);
\draw[dashed] (12,-2) -- node[black,above,yshift=-0.6cm,xshift=-0.2cm] {$H_i$} (14,-2.2);
\end{tikzpicture}
\hspace{5mm}
\begin{tikzpicture}[thick,scale=0.8]
\draw[particle] (10,0.2) -- node[black,above,xshift=-0.9cm,yshift=0.1cm] {$e^-$} (12,0);
\draw[antiparticle] (10,-3.2) -- node[black,above,yshift=-0.7cm,xshift=-0.9cm] {$e^+$} (12,-3.0);
\draw[particle] (12,0) -- node[black,above,xshift=1.1cm,yshift=0.1cm] {$e^-$} (14,0.2);
\draw[antiparticle] (12,-3.0) -- node[black,above,yshift=-0.7cm,xshift=1.1cm] {$e^+$} (14,-3.2);
\draw[photon] (12,0) -- node[black,above,yshift=-0.2cm,xshift=-0.4cm] {$Z$} (12,-1.5);
\draw[photon] (12,-1.5) -- node[black,above,yshift=-0.4cm,xshift=-0.4cm] {$Z$} (12,-3.0);
\draw[dashed] (12,-1.5) -- node[black,above,yshift=-0.1cm,xshift=-0.2cm] {$H_i$} (14,-0.8);
\draw[dashed] (12,-1.5) -- node[black,above,yshift=-0.6cm,xshift=-0.2cm] {$H_i$} (14,-2.2);
\end{tikzpicture}
\hspace{5mm}
\begin{tikzpicture}[thick,scale=0.8]
\draw[particle] (10,0.2) -- node[black,above,xshift=-0.9cm,yshift=0.1cm] {$e^-$} (12,0);
\draw[antiparticle] (10,-3.2) -- node[black,above,yshift=-0.7cm,xshift=-0.9cm] {$e^+$} (12,-3.0);
\draw[particle] (12,0) -- node[black,above,xshift=1.1cm,yshift=0.1cm] {$e^-$} (14,0.2);
\draw[antiparticle] (12,-3.0) -- node[black,above,yshift=-0.7cm,xshift=1.1cm] {$e^+$} (14,-3.2);
\draw[photon] (12,0) -- node[black,above,yshift=-0.2cm,xshift=-0.4cm] {$Z$} (12,-1.5);
\draw[photon] (12,-1.5) -- node[black,above,yshift=-0.4cm,xshift=-0.4cm] {$Z$} (12,-3.0);
\draw[dashed] (12,-1.5) -- node[black,above,yshift=-0cm,xshift=-0cm] {$h$} (13,-1.5);
\draw[dashed] (13,-1.5) -- node[black,above,yshift=-0.1cm,xshift=-0.2cm] {$H_i$} (15,-0.8);
\draw[dashed] (13,-1.5) -- node[black,above,yshift=-0.6cm,xshift=-0.2cm] {$H_i$} (15,-2.2);
\end{tikzpicture}
\caption{The $t$-channel diagrams leading to the $2\ell + \Et$ final state mediated by the $Z, h$ and $A_i$ propagators.}
\label{fig:Et2l-ILC-t}
\end{figure}
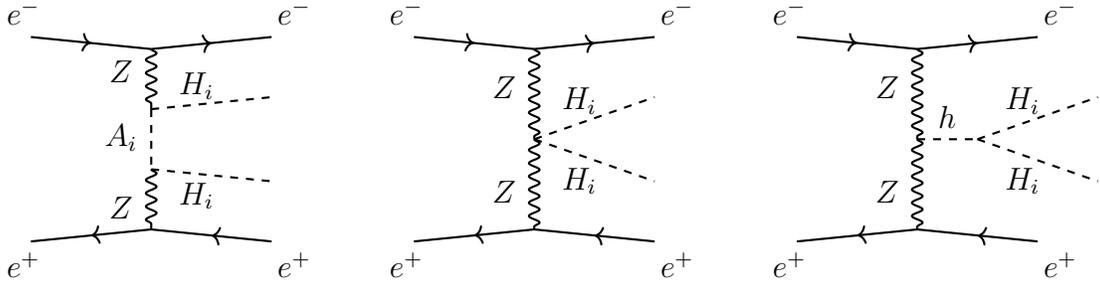

In Figs.~\ref{fig:Et2l-ILC-s}--\ref{fig:Et2l-ILC-t}, we show the dominant $s$-channel and $t$-channel diagrams which contribute to the $e^+ e^- \rightarrow  \ell^+ \ell^- + \mathrm{\rm DM}\, \mathrm{\rm DM}$ process,  where DM is represented by $H_i$. The diagrams for $A_i$ as DM are identical with the interchange of $H_i \leftrightarrow A_i$. However, do recall that the Feynman rules are different (in the Higgs-DM coupling) in the two sets of graphs.

\section{Theoretical and Experimental Constraints}
\label{sec:constraints}

The parameters of the potential are subject to a number of theoretical and experimental constraints (described in detail  in 
\cite{Hernandez-Sanchez:2020aop}). Below, we summarise the constraints imposed on the model to ensure that our BPs are phenomenologically viable.

\begin{itemize}
\item 
\textbf{Stability of the potential}

For the potential to remain bounded from below (i.e., to ensure a stable vacuum), the following constraints must be satisfied \cite{Grzadkowski:2009bt}:
\bea
&& \lambda_{ii}>0, \quad i =1,2,3, \label{positivity1} \\
&& \lambda_x > - 2 \sqrt{\lambda_{11} \lambda_{22}}, \quad \lambda_y > - 2 \sqrt{\lambda_{11} \lambda_{33}}, \quad \lambda_z > - 2 \sqrt{\lambda_{22} \lambda_{33}}, \label{positivity2}\\[1mm]
&&\left\lbrace  \begin{array}{l} 
\sqrt{\lambda_{33}} \lambda_x + \sqrt{\lambda_{11}} \lambda_z + \sqrt{\lambda_{22}} \lambda_y \geq 0\\
\quad \text{or}\\[1mm]
\lambda_{33} \lambda_x^2 + \lambda_{11} \lambda_z^2 + \lambda_{22} \lambda_y^2 -\lambda_{11} \lambda_{22} \lambda_{33} - 2 \lambda_x \lambda_y \lambda_z < 0,
\end{array}\right.
\label{positivity3}
\eea
where 
\bea
\lambda_x = \lambda_{12} + \min(0, \lambda_{12}' - 2|\lambda_1|),\\[1mm]
\lambda_y = \lambda_{31} + \min(0, \lambda_{31}' - 2|\lambda_3|),\\[1mm]
\lambda_z = \lambda_{23} + \min(0, \lambda_{23}' - 2|\lambda_2|).
\eea
As discussed in \cite{Faro:2019vcd}, these conditions are sufficient but not necessary. It is indeed possible to find examples within this model where the potential is bounded from below, even if the criteria in Eqs.~(\ref{positivity1})--(\ref{positivity3}) are not met. However, in this work, we do not explore such parameter regions.


\item 
\textbf{Global minimum condition}

For the point $(0,0,v)$ to be a local minimum, all mass-squared terms must be positive. To ensure it is also the global minimum, its energy must be lower than the energy of any other coexisting minima (see detailed discussion in \cite{Hernandez-Sanchez:2020aop}), therefore, the following conditions are required:
\bea
 \text{Local minimum if:}&& \left\lbrace \begin{array}{l}
v^2 = {\mu_3^2}/{\lambda_{33}} > 0,\\[1.5mm]
\min(\Lambda_2, \bar \Lambda_2) > {\mu_2^2}/{v^2},\\[1.5mm]
\min(\Lambda_3, \bar \Lambda_3) > {\mu_1^2}/{v^2}.
\end{array} \right.
\hspace{3cm}
\label{inert-loc}
\\[2mm]
\text{Global minimum if, additionally:}&&  \left\lbrace \begin{array}{l}
{\mu_3^2}/{\sqrt{\lambda_{33}}} > {\mu_2^2}/{\sqrt{\lambda_{22}}} ,\\[1.5mm]
{\mu_3^2}/{\sqrt{\lambda_{33}}} > {\mu_1^2}/{\sqrt{\lambda_{11}}} .
\end{array} \right.
\hspace{3cm}
\label{inert-glob}
\eea


\item 
\textbf{Perturbative unitarity} 

We impose the condition that the scalar $2 \to 2$ scattering matrix remains unitary, meaning that the absolute values of all eigenvalues of this matrix for Goldstone bosons, Higgs and dark states with given hypercharge and isospin must be less than $8 \pi$. Additionally, all quartic scalar couplings are required to stay within the perturbative regime, {i.e.}, $\lambda_i \leq 4 \pi$.


\item 
\textbf{EW Precision Observables (EWPOs)} 

We demand a 2$\sigma$, i.e., 95\% Confidence Level (CL) agreement with EWPOs which are  parametrised through the EW oblique parameters $S,T,U$. Assuming an SM Higgs boson mass of $m_h$ = 125 GeV, the central values of the oblique parameters are given by~\cite{Baak:2014ora}:
\be 
\hat{S} = 0.05 \pm 0.11 ,\qquad \hat{T} = 0.09 \pm 0.13, \qquad \hat{U}=0.01\pm 0.11.
\label{eq:ewpt}
\ee
In the I(1+1)HDM, these constraints impose a strict order on the masses of the inert particles, with two neutral dark particles being lighter than the charged dark particle. Furthermore, mass splitting between the heavier neutral sate and the charged state  is limited to roughly 50 GeV. However, in the case of a $Z_2 \times Z_2'$-symmetric I(2+1)HDM,  these conclusions are no longer applicable. Cancellations between contributions to the $S,T,U$ parameters from the two generations of dark particles may lead to a different mass orderings, where either of $A_i$ or $H^\pm_i$ is the heaviest, as well as to an increased mass splittings between these particles (for a detailed discussion, see \cite{Hernandez-Sanchez:2020aop}).


\item 
\textbf{Collider searches for new physics}

The introduction of additional scalar fields, particularly if they are sufficiently light, can impact the properties of SM particles, such as their decay modes and widths. To prevent decays of the EW gauge bosons into these new scalars, we impose the following constraints:
\be 
\label{eq:gwgz}
m_{H_i} + m_{H^\pm_i} \,\geq\, m_W^\pm,~~ m_{A_i} + m_{H^\pm_i} \,\geq\, m_W^\pm,~~
m_{H_i} + m_{A_i} \,\geq\, m_Z,\,~~
2\,m_{H_i^\pm} \,\geq\, m_Z.
\ee
Additionally, we apply limits from LEP 2 searches for supersymmetric particles, re-interpreted for the I(1+1)HDM, to exclude mass regions where the following conditions hold simultaneously \cite{Lundstrom:2008ai} ($i=1,2$):
\be 
\label{eq:leprec}
m_{A_i} \,\leq\, 100\,\GeV,\,~~
m_{H_i} \,\leq\, 80\,\GeV,\,~~
\Delta m = |m_{A_i} - m_{H_i}| \,\geq\, 8\,\GeV,
\ee
since this would generate a visible di-jet or di-lepton signal.

Furthermore, the model must be consistent with null results from LHC searches for additional scalars. As discussed in \cite{Hernandez-Sanchez:2020aop}, current LHC searches for multi-lepton final states with missing transverse energy, $\Et$, are generally not sensitive enough to test the parameter space of this model. This is mainly due to the relatively high $\Et$ cut used in experimental analyses, which reduces the sensitivity to the viable parameter space of the I(2+1)HDM framework. Moreover, since the new charged particles are inert and do not couple to fermions, constraints typically applicable in the 2HDM framework, such as those from $b \to s \gamma$ processes, do not apply here.
 

\item 
\textbf{Charged scalar mass and lifetime}

We adopt a model-independent lower bound on the masses of all charged states, specifically, $m_{H^\pm_i} > 70\,\GeV$ ($i=1,2$) \cite{Pierce:2007ut}. In addition, we do not consider scenarios involving potentially long-lived charged particles. Following \cite{Heisig:2018kfq}, we impose an upper limit on the lifetime of charged states, requiring $\tau \leq 10^{-7} \, \text{s}$.


\item 
\textbf{Higgs mass and signal strengths} 

The combined measurement of the Higgs boson mass by ATLAS and CMS is given as~\cite{ATLAS:2015yey}:
\be
m_h = 125.09 \pm 0.21 \, \text{(stat.)} \pm 0.11 \, \text{(syst.)} \; \GeV.
\ee
The observed Higgs particle at the LHC matches the SM predictions very well. By construction, the $h$ state in Eq.~(\ref{vac-inert}) behaves like the SM Higgs boson, with its couplings to gluons, massive gauge bosons and fermions being identical to those of the SM-Higgs (at tree level).

The total Higgs width can be altered through additional decay channels into light inert scalars, $S$, leading to a contribution from the $h \to SS$ decay when $m_S \leq m_h / 2$. Moreover, there can be modifications to existing SM decay channels, especially the $h \to \gamma\gamma$ process. In this work, we use an upper bound on the Higgs total decay width as reported in~\cite{Sirunyan:2019twz}:
\be
\Gamma_{\rm tot} \leq 9.1 \; \text{MeV}.
\ee
The partial decay width $\Gamma(h \to \gamma\gamma)$ is modified due to the presence of two charged inert scalars. We ensure compatibility with the ATLAS and CMS limit on this signal strength, as given in~\cite{Khachatryan:2016vau},
\be
\mu_{\gamma \gamma} = 1.14^{+0.19}_{-0.018},
\ee
satisfying a 2$\sigma$ agreement with experimental observations.

The most recent constraints on the Higgs invisible decays from CMS and ATLAS are~\cite{Sirunyan:2018owy, Aaboud:2019rtt}:
\be
\text{BR}(h \to \text{ inv.}) < 0.19 \; \text{(CMS)}, \quad 0.26 \; \text{(ATLAS)}.
\ee
These bounds place strong restrictions on the Higgs-inert couplings, particularly for light inert scalars.


\item 
\textbf{DM constraints}

The total relic density is the sum of the individual contributions from both DM components, $H_1$ and $H_2$ for BP1 (or $H_1$ and $A_2$ in case of BP2), and can be expressed as:
\be 
\label{planck-relic}
\Omega_{T}h^2 = \Omega_{{H_1}}h^2 + \Omega_{{H_2(A_2)}}h^2 \, ,
\ee
which is constrained by the Planck satellite data \cite{Aghanim:2018eyx} to be:
\be
\Omega_{\text{\rm DM}}h^2 = 0.1200 \pm 0.0012.
\label{PLANCK_lim}
\ee

The current most stringent upper bounds on the Spin-Independent (SI) DM-nucleon scattering cross section, $\sigma_{{\rm DM}-N}$, are set by the XENON1T and PandaX-4T experiments, relevant for a wide range of DM masses \cite{Aprile:2018dbl, PandaX-4T:2021bab}. 

For indirect detection, the tightest constraints on light DM candidates annihilating into $b\bar{b}$ or $\tau^+\tau^-$ final states come from the Fermi-LAT satellite, excluding the canonical annihilation cross section 
$\langle \sigma v \rangle \approx 3 \times 10^{-26} \, \mathrm{cm}^3/\mathrm{s}$ for $m_{\rm DM} \lesssim 100 \, \text{GeV}$ \cite{Ackermann:2015zua}. 
For heavier DM candidates, the PAMELA and Fermi-LAT experiments set similar limits, with $ \langle \sigma v \rangle \approx 10^{-25} \, \mathrm{cm}^3/\mathrm{s}$ for $m_{\rm DM} \approx 200 \, \text{GeV}$ in the $b\bar{b}$, $\tau^+\tau^-$ or $W^+W^-$ annihilation channels \cite{Cirelli:2013hv}.

\end{itemize}

\section{Collider Analysis and Distributions}\label{sec:collider}

Recall that our signal is $e^+ e^- \rightarrow  \ell^+ \ell^- +  \mathrm{\rm DM}\, \mathrm{\rm DM} $, where DM will escape the detector resulting in missing energy in the final state along with two opposite sign leptons. As mentioned already, here,  we consider both electrons and muons in the final state.

To see the differences coming from two different BPs in an $e^+ e^-$ collider, we generate our signal events for, e.g., 1 TeV centre-of-mass energy. In this case we are trying to follow the experimental set-up of the International Linear Collider (ILC)~\cite{Behnke:2013xla,ILC:2013jhg,Adolphsen:2013jya,Adolphsen:2013kya,Behnke:2013lya}  and also  consider that our $e^-$ and $e^+$ beams are 80\% and 30\% polarised, respectively. For detector level analysis we have used the dedicated \texttt{Delphes} card forthe  ILC. 

To conduct a comprehensive analysis at the detector level, we focus on the distribution patterns of several key observables for both signals, as shown in Tab.~\ref{tab:observables}.
We have defined $m_{transverse}$ as
\be 
m_{transverse} = \sqrt{\left(\sqrt{{m^2}_{\ell_1,\ell_2}+{P^2_T}_{\ell_1,\ell_2}}+\Et \right)^2-{P^2_T}_{\ell_1,\,\ell_2,\,\Et}},
\label{MT}
\ee
where ${P_T}_{\ell_1,\ell_2}$ is the vector sum of the transverse momentum of two lepton system with highest momentum, and ${P_T}_{\ell_1,\ell_2,\Et}$ is the vector sum of the transverse momentum of two (leading) leptons along with missing transverse energy ($\Et$).
The observable $P_{\theta}$ is defined as
\be 
P_{\theta} = 
\frac{ \left| E_{miss}-E_{\ell_1,\ell_2} \right| }{E_{miss}+E_{\ell_1,\ell_2}},
\label{PTheta}
\ee
where $E_{\ell_1,\ell_2}$ is the sum of the energy of two (leading) leptons. This is an observable which carries the information of the fraction of energy imbalance between two leptons and the missing energy system.

\begin{table}[htpb!]
\begin{footnotesize}
\begin{tabular}{|l||p{14cm}|} 
\hline
\textbf{Observable} & \textbf{Description} \\
\hline\hline
${P_T}_{\ell_1}$ & Transverse momentum of leading lepton (the lepton that carries highest momentum)\\ 
\hline
${P_T}_{\ell_2}$ & Transverse momentum of sub-leading lepton (the lepton carries second highest momentum)\\ 
\hline
$\Et$ & Missing transverse momentum\\
\hline
$E_{miss}$ & Missing energy\\
\hline
$m_{transverse}$ 
& Transverse mass of final state including two lepton and missing energy\\
\hline
$m_{\ell_1,\ell_2,E_{miss}}$& Invariant mass of two leading leptons and missing energy system\\
\hline
$\Delta \eta_{\ell_1,\ell_2}$& Difference of pseudo-rapidity between two leading leptons with highest momentum\\
\hline
$\Delta R_{\ell_1,\ell_2}$& Radial distance between two leading leptons with highest momentum\\
\hline
$\Delta \phi_{\ell_1,\ell_2}$& Difference of azimuthal angle between two leading leptons with highest momentum\\ 
\hline
$P_{\theta}$& Energy imbalance between missing energy and two leading lepton system\\
\hline
\end{tabular}
\end{footnotesize}
\caption{Final state observables used to show the distributions of signal coming from BP1 and BP2.}
\label{tab:observables}
\end{table}

\begin{figure}[h!]
\centering
\includegraphics[scale=0.9]{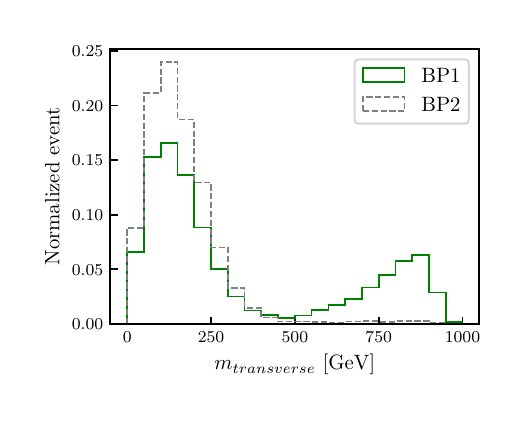} ~
\includegraphics[scale=0.9]{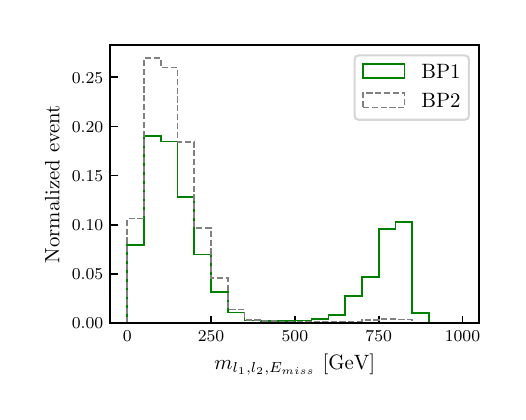}
\vspace{-5mm}
\caption{Normalised distribution of the transverse mass of two lepton and missing energy final state, $m_{transverse}$ (left) and the invariant mass of two leading leptons and missing energy, $m_{\ell_1,\ell_2,E_{miss}}$ (right) at a 1 TeV ILC where $e^-$ and $e^+$ are 80\% and 30\% polarised, respectively, for BP1 and BP2 after detector simulation.}
\label{fig:mtransverse-ml1l2miss}
\end{figure}

\begin{figure}[h!]
\centering
\includegraphics[scale=0.9]{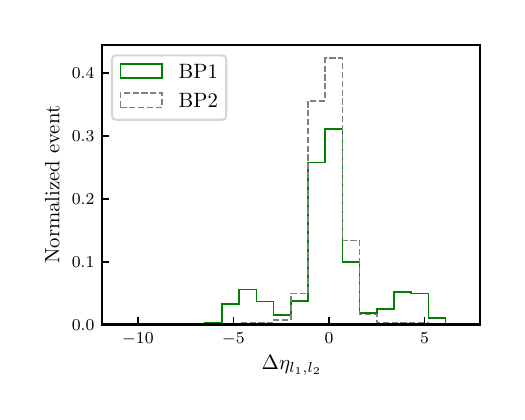} ~
\includegraphics[scale=0.9]{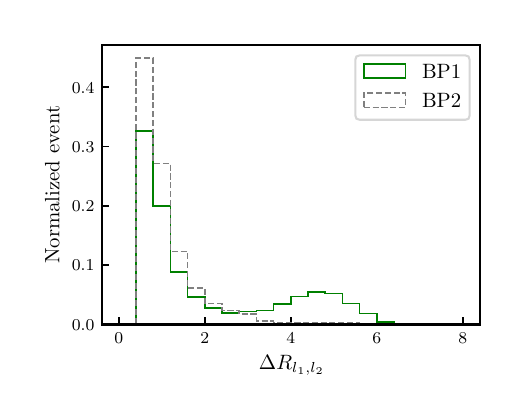}
\vspace{-5mm}
\caption{Normalised distribution of the difference of pseudo-rapidity of two leading leptons $\Delta\eta_{\ell_1,\ell_2}$ on the left, and the radial separation between two leading lepton $\Delta R_{\ell_1,\ell_2}$ on the right at a 1 TeV ILC where $e^-$ and $e^+$ are 80\% and 30\% polarised, respectively,   for BP1 and BP2 after detector simulation.}
\label{fig:detal1l2-drl1l2}
\end{figure}

\begin{figure}[h!]
\centering
\includegraphics[scale=0.9]{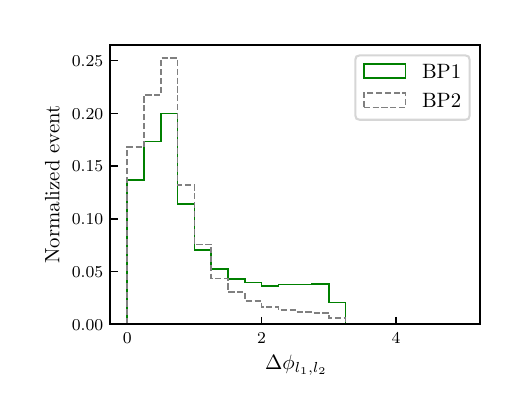} ~
\includegraphics[scale=0.9]{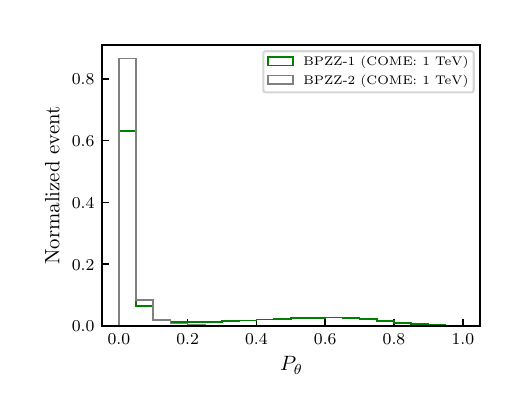}
\vspace{-5mm}
\caption{Normalised distribution of the angular separation between two leading lepton in azimuthal plane $\Delta \Phi_{\ell_1,\ell_2}$ on the left and the energy imbalance between missing energy and two leading lepton system $P_{\theta}$ on the right at a 1 TeV ILC where $e^-$ and $e^+$ are 80\% and 30\% polarised, respectively,   for BP1 and BP2 after detector simulation.}
\label{fig:dphil1l2-ptheta}
\end{figure}

We analyse the distribution profiles of all  these final-state distributions  to identify observables which can distinguish between the two BPs, {i.e.}, whether the two DM components have opposite or identical CP.
In Figs.~\ref{fig:mtransverse-ml1l2miss}--\ref{fig:dphil1l2-ptheta}, we show the distributions for an $e^+ e^-$ linear collider operating at a center-of-mass energy of 1 TeV. 
In all plots, the solid distributions represent BP1 while the dashed ones  correspond to BP2.

In Fig.~\ref{fig:mtransverse-ml1l2miss}, the panel on the left shows the $m_{transverse}$ observable where a clear distinction between the two BPs in visible, particularly in the higher transverse mass region. In BP1, where the two DM components have the same CP, the events tend to favour higher-energy final states involving leptons. In contrast, in BP2, where the two DM components have opposite CP, the distribution shows rather  different characteristics.
The right panel shows the distribution of the invariant mass of the final state, $m_{\ell_1,\ell_2,E_{miss}}$. Here, BP2 predominantly favours the low-mass region whereas BP1 produces significant contributions in both high and low mass regions. This pattern reflects the influence of interference among various diagrams, particularly due to the coupling of the SM-like Higgs with the second DM candidate as well as the momentum-dependent contributions from $t$-channel processes. 

Fig.~\ref{fig:detal1l2-drl1l2} shows the distributions of $\Delta \eta_{\ell_1,\ell_2}$ (on the left), where it is evident that BP1 features many events with two highly boosted forward leptons separated by large angular differences in the transverse plane while, in contrast, BP2 displays leptons that remain close together. This distinction is partly due to the fact that BP1 has a preference for high $P_T$ leptons, whereas BP2 favours softer  leptons. The differences observed in $\Delta \eta_{\ell_1,\ell_2}$ also affect the angular separation of the two final state leptons, $\Delta R_{\ell_1,\ell_2}$ (on the right). For BP1, we often find leptons that are widely separated and high in energy. These effects are strongly influenced by the $t$-channel diagrams shown in Fig.~\ref{fig:Et2l-ILC-t}, where we see two highly boosted forward electrons with missing energy contributions in the central region. 

In Fig.~\ref{fig:dphil1l2-ptheta}, on the left, we show $\Delta \phi_{\ell_1,\ell_2}$, where there are more events with large angular separation in BP1 compared to BP2. On the right, we show the observable $P_{\theta}$, which measures the fraction of energy imbalance between the lepton systems and DM particles, the latter  reflected the missing transverse energy. The distribution of $P_{\theta}$ reveals that, in BP1, there are events where the majority of the final state energy is carried by either the lepton system or the DM particles, rather than being equally shared between them. This pattern is absent in BP2, where the energy is consistently shared equally between the lepton system and the DM particles.
 
 \subsection{The Importance of the $t$-Channel Diagrams}\label{sec:$t$-channel}
 
\begin{figure}[h!]
\includegraphics[scale=0.9]{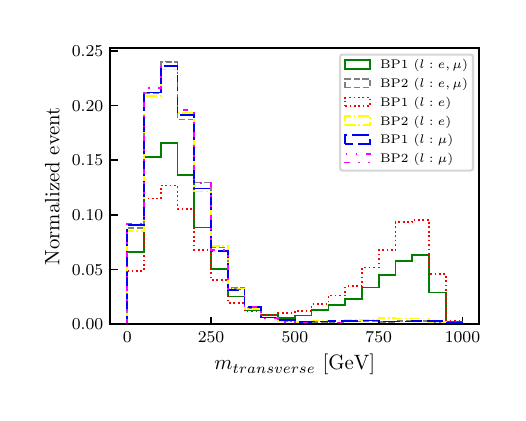}~
\includegraphics[scale=0.9]{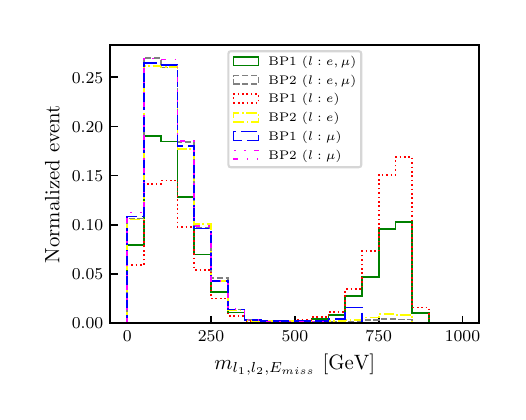}
\caption{Normalised distribution of the transverse mass of two lepton and missing energy final state, $m_{transverse}$ (left) and the invariant mass of two leading leptons and missing energy, $m_{\ell_1,\ell_2,E_{miss}}$ (right) at a 1 TeV ILC where $e^-$ and $e^+$ are 80\% and 30\% polarised, respectively, for BP1 and BP2 after detector simulation.}
\label{fig:mtransverse-ml1l2miss-emu}
\end{figure}

Recall that our `smoking-gun' signal is found at an ILC through the $e^+ e^- \rightarrow \ell^+ \ell^- + \mathrm{\rm DM}\, \mathrm{\rm DM} $ process,
where $\ell=e,\mu$. However,  
note that the $t$-channel diagrams in Figs.~\ref{fig:Et2l-ILC-s}--\ref{fig:Et2l-ILC-t} would be zero for $e^+ e^- \rightarrow  \mu^+ \mu^- + \mathrm{\rm DM}\, \mathrm{\rm DM} $ and only $s$-channel diagrams would contribute. This observation enables us to 
 illustrate that the aforementioned interference effect is primarily occurring between the $t$- and $s$-channel diagrams. To this end,  we now plot the aforementioned observables for the $\mu^+ \mu^-$ final state and compare these   with the yield of the $e^+ e^-$ final state. The result is shown in Figs.~\ref{fig:mtransverse-ml1l2miss-emu}--\ref{fig:detal1l2-drl1l2-emu}. In all plots, one can see that the difference in the distributions of BP1 versus BP2 is a lot less pronounced when only muons are in the final state, {i.e.}, the $t$-channel diagram from Fig.~\ref{fig:Et2l-ILC-t} are absent. 
This also explains why the difference between the two BPs seems larger when only electrons are in the final state in comparison to when both electrons and muons are allowed in the final state, so that one would be  best placed in experimentally pursuing  specifically the $\ell=e$ case. Finally, notice that we have shown distributions normalised to the same area, to emphasise the differences in shape, yet, the cross sections reported in Tab.~\ref{tab:BPs} (in the caption)  illustrate that both CP  hypotheses can be tested over the same time scale at a 1 TeV ILC, however, we refrain here from estimating the actual sensitivity of this machine to such processes, crucially, in presence of backgrounds and, hence, of a dedicated signal-to-background analysis: in fact, we leave this to future endeavours.
 
\begin{figure}[h!]
\includegraphics[width=8.5cm, height=6.5cm]{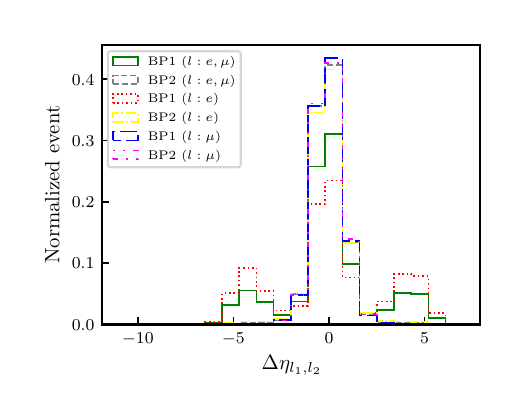}
\includegraphics[width=8.5cm, height=6.5cm]{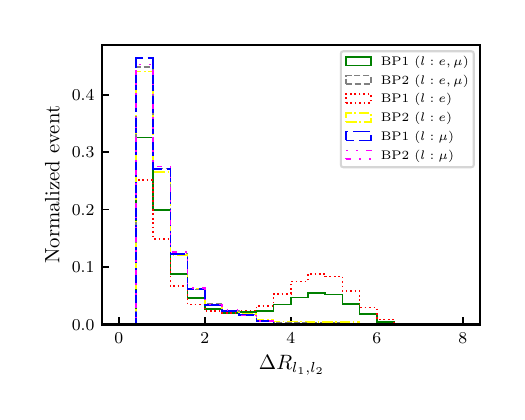}
\caption{Normalised distribution of the difference of pseudorapidity of two leading leptons, $\eta_{\ell_1,\ell_2}$ (left) and the radial separation between two leading lepton,$\Delta R_{\ell_1,\ell_2}$ (right) at a 1 TeV ILC where $e^-$ and $e^+$ are 80\% and 30\% polarised, respectively,   for BP1 and BP2 after detector simulation.}
\label{fig:detal1l2-drl1l2-emu}
\end{figure}

\begin{figure}[h!]
\includegraphics[width=8.5cm, height=6.5cm]{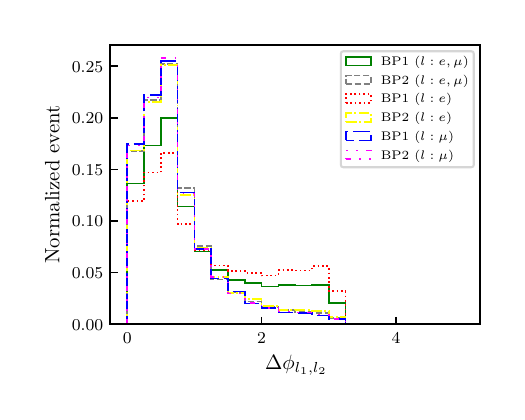}
\includegraphics[width=8.5cm, height=6.5cm]{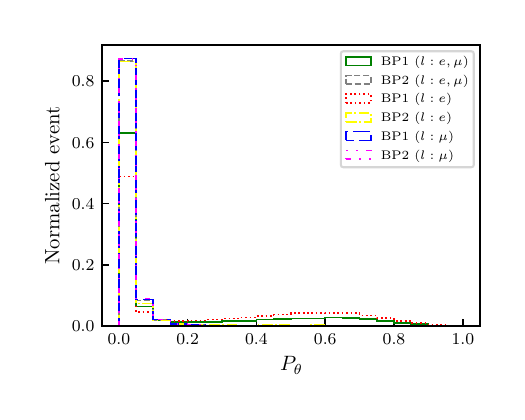}
\caption{Normalised distribution of the angular separation between two leading lepton in azimuthal plane, $\Delta \Phi_{\ell_1,\ell_2}$ (left) and the Energy imbalance between missing energy and two leading lepton system, $P_{\theta}$ (right) at a 1 TeV ILC where $e^-$ and $e^+$ are 80\% and 30\% polarised, respectively,   for BP1 and BP2 after detector simulation.}
\label{fig:detal1l2-drl1l2-emu}
\end{figure}

\section{Conclusions}
\label{sec:conclusion}

It is known that in the IDM (also referred to as the I(1+1)HDM), one cannot identify the CP of the DM candidate. Herein, the lightest spin-0 particle from the inert (or dark) sector is automatically the DM candidate and could be either CP-even or CP-odd. However, extensions of this BSM construct,  such as the I(2+1)HDM,  allow for more complex DM scenarios. In this paper, we have analysed a $Z_2 \times Z_2$-symmetric I(2+1)HDM that is known to provide a two-component DM setup, one from each family of dark particles, where the first family consists of $H_1,A_1, H^\pm_1$ and the second family consists of  $H_2,A_2, H^\pm_2$, i.e., with the labels 1 and 2 referring to each of the two inert doublets. Specifically, 
the lightest neutral field from each family provides one component of DM. In this construct, the relative CP state of the two DM candidates can be such that both the following possibilities are allowed:   the two DM  components can have the same or opposite CP quantum numbers.

Therefore, in this paper, by leveraging phenomenological aspects of this model that we have explored in previous literature, we have set out to understand whether there could be collider signals revealing which of the two above CP hypotheses is realised in Nature. By assuming that both CP scenarios involve $H_1$ as one of the two DM particles (although similar results can be obtained if $A_1$ were chosen), we have looked at whether one could disentangle the second DM component between the $H_2$ (same CP) and $A_2$ (opposite CP) assumptions.     In order to do so, we have devised two BPs in agreement with all theoretical and experimental bounds: as intimated, in one scenario the two DM components have the same CP while in the other  the two DM components have opposite CP.  In both cases, the coupling carrying the hallmark of the relative CP status between the two DM particles, i.e., the one between the  SM-like Higgs boson and each DM state, has the same size but not the same sign, as the latter depends on whether one is dealing with $H_2$ or $A_2$, thereby inducing in turn  constructive or destructive interferences.  

We have thus shown that an ideal experimental setup to accomplish the above goal is constituted by an $e^+ e^-$  linear collider, ideally exploiting both beam polarisation. By adopting the ILC prototype with a 1 TeV energy, we have   identified the $e^+ e^- \rightarrow \ell^+ \ell^- + \mathrm{\rm DM}\, \mathrm{\rm DM}$ process ($\ell = e, \mu$) as our best signal. In particular, notice  that, in our first benchmark scenario (same CP), this signal receives contributions from the $e^+ e^- \rightarrow \ell^+ \ell^- + H_1H_1$ and $e^+ e^- \rightarrow \ell^+ \ell^- + H_2H_2$ processes while, in our second benchmark scenario, this signal receives contributions from the $e^+ e^- \rightarrow \ell^+ \ell^- + H_1H_1$ and $e^+ e^- \rightarrow \ell^+ \ell^- + A_2A_2$ processes.

By studying several kinematic observables defined for the ensuing $e^+e^-\to \ell^+\ell^- + \Et$ signature, we have proven that the two above hypotheses can be separated on a similar timescale 
 as they present noticeably different shapes and comparable cross sections. Specifically, we have identified that the leading effect responsible for such differences is due to interference between two specific sets of topologies, an $s$- and a $t$-channel one. Furthermore, as the latter is absent for the case $\ell=\mu$, we have emphasised the crucial role of the $\ell=e$ case, which is the one that should ideally  be pursued in future phenomenological studies, also accounting for background effects as well as a dedicated signal-to-background analysis. Altogether, we strongly advocate the latter, as there is clear potential for two-component DM characterisation in our approach:  although one cannot identify the CP state of each DM candidate, it is possible to access the relative CP status of the two DM components.

\subsubsection*{Acknowledgement}
VK and AD acknowledge financial support from the Science Foundation Ireland Grant 21/PATH-S/9475 (MOREHIGGS) under the SFI-IRC Pathway Programme. JH-S acknowledges the support by SNI-CONAHCYT and VIEP-BUAP under the grant "Higgs and dark matter physics".
SM acknowledges support from the STFC Consolidated Grant ST/L000296/1 and is partially financed through the NExT Institute.
TS acknowledges finantial support from the JSPS KAKENHI Grant Number 20H00160.


\begin{thebibliography}{99}

\bibitem{Aad:2012tfa}
G.~Aad \textit{et al.} [ATLAS],
Phys. Lett. B \textbf{716}, 1-29 (2012)
doi:10.1016/j.physletb.2012.08.020
[arXiv:1207.7214 [hep-ex]].

\bibitem{Chatrchyan:2012ufa}
S.~Chatrchyan \textit{et al.} [CMS],
Phys. Lett. B \textbf{716}, 30-61 (2012)
doi:10.1016/j.physletb.2012.08.021
[arXiv:1207.7235 [hep-ex]].

\bibitem{Ade:2015xua}
P.~A.~R.~Ade \textit{et al.} [Planck],
Astron. Astrophys. \textbf{594}, A13 (2016)
doi:10.1051/0004-6361/201525830
[arXiv:1502.01589 [astro-ph.CO]].

\bibitem{Jungman:1995df}
G.~Jungman, M.~Kamionkowski and K.~Griest,
Phys. Rept. \textbf{267}, 195-373 (1996)
doi:10.1016/0370-1573(95)00058-5
[arXiv:hep-ph/9506380 [hep-ph]].

\bibitem{Bertone:2004pz}
G.~Bertone, D.~Hooper and J.~Silk,
Phys. Rept. \textbf{405}, 279-390 (2005)
doi:10.1016/j.physrep.2004.08.031
[arXiv:hep-ph/0404175 [hep-ph]].

\bibitem{Bergstrom:2000pn}
L.~Bergstr\"om,
Rept. Prog. Phys. \textbf{63}, 793 (2000)
doi:10.1088/0034-4885/63/5/2r3
[arXiv:hep-ph/0002126 [hep-ph]].

\bibitem{Nilles:1983ge}
H.~P.~Nilles,
Phys. Rept. \textbf{110}, 1-162 (1984)
doi:10.1016/0370-1573(84)90008-5

\bibitem{Haber:1984rc}
H.~E.~Haber and G.~L.~Kane,
Phys. Rept. \textbf{117}, 75-263 (1985)
doi:10.1016/0370-1573(85)90051-1

\bibitem{Cheng:2002ej}
H.~C.~Cheng, J.~L.~Feng and K.~T.~Matchev,
Phys. Rev. Lett. \textbf{89}, 211301 (2002)
doi:10.1103/PhysRevLett.89.211301
[arXiv:hep-ph/0207125 [hep-ph]].

\bibitem{Servant:2002aq}
G.~Servant and T.~M.~P.~Tait,
Nucl. Phys. B \textbf{650}, 391-419 (2003)
doi:10.1016/S0550-3213(02)01012-X
[arXiv:hep-ph/0206071 [hep-ph]].

\bibitem{McDonald:1993ex}
J.~McDonald,
Phys. Rev. D \textbf{50}, 3637-3649 (1994)
doi:10.1103/PhysRevD.50.3637
[arXiv:hep-ph/0702143 [hep-ph]].

\bibitem{Burgess:2000yq}
C.~P.~Burgess, M.~Pospelov and T.~ter Veldhuis,
Nucl. Phys. B \textbf{619}, 709-728 (2001)
doi:10.1016/S0550-3213(01)00513-2
[arXiv:hep-ph/0011335 [hep-ph]].

\bibitem{Deshpande:1977rw}
N.~G.~Deshpande and E.~Ma,
Phys. Rev. D \textbf{18}, 2574 (1978)
doi:10.1103/PhysRevD.18.2574

\bibitem{Ma:2006km}
E.~Ma,
Phys. Rev. D \textbf{73}, 077301 (2006)
doi:10.1103/PhysRevD.73.077301
[arXiv:hep-ph/0601225 [hep-ph]].

\bibitem{Belanger:2012zr}
G.~Belanger, K.~Kannike, A.~Pukhov and M.~Raidal,
JCAP \textbf{01}, 022 (2013)
doi:10.1088/1475-7516/2013/01/022
[arXiv:1211.1014 [hep-ph]].

\bibitem{Barbieri:2006dq}
R.~Barbieri, L.~J.~Hall and V.~S.~Rychkov,
Phys. Rev. D \textbf{74}, 015007 (2006)
doi:10.1103/PhysRevD.74.015007
[arXiv:hep-ph/0603188 [hep-ph]].

\bibitem{LopezHonorez:2006gr}
L.~Lopez Honorez, E.~Nezri, J.~F.~Oliver and M.~H.~G.~Tytgat,
JCAP \textbf{02}, 028 (2007)
doi:10.1088/1475-7516/2007/02/028
[arXiv:hep-ph/0612275 [hep-ph]].

\bibitem{Ivanov:2012hc}
I.~P.~Ivanov and V.~Keus,
Phys. Rev. D \textbf{86}, 016004 (2012)
doi:10.1103/PhysRevD.86.016004
[arXiv:1203.3426 [hep-ph]].

\bibitem{Patt:2006fw}
B.~Patt and F.~Wilczek,
[arXiv:hep-ph/0605188 [hep-ph]].

\bibitem{Chu:2011be}
X.~Chu, T.~Hambye and M.~H.~G.~Tytgat,
JCAP \textbf{05}, 034 (2012)
doi:10.1088/1475-7516/2012/05/034
[arXiv:1112.0493 [hep-ph]].

\bibitem{Queiroz:2014yna}
F.~S.~Queiroz and K.~Sinha,
Phys. Lett. B \textbf{735}, 69-74 (2014)
doi:10.1016/j.physletb.2014.06.016
[arXiv:1404.1400 [hep-ph]].

\bibitem{Mambrini:2011ik}
Y.~Mambrini,
Phys. Rev. D \textbf{84}, 115017 (2011)
doi:10.1103/PhysRevD.84.115017
[arXiv:1108.0671 [hep-ph]].

\bibitem{Djouadi:2011aa}
A.~Djouadi, O.~Lebedev, Y.~Mambrini and J.~Quevillon,
Phys. Lett. B \textbf{709}, 65-69 (2012)
doi:10.1016/j.physletb.2012.01.062
[arXiv:1112.3299 [hep-ph]].

\bibitem{Djouadi:2012zc}
A.~Djouadi, A.~Falkowski, Y.~Mambrini and J.~Quevillon,
Eur. Phys. J. C \textbf{73}, no.6, 2455 (2013)
doi:10.1140/epjc/s10052-013-2455-1
[arXiv:1205.3169 [hep-ph]].

\bibitem{Belanger:2012vp}
G.~Belanger, K.~Kannike, A.~Pukhov and M.~Raidal,
JCAP \textbf{04}, 010 (2012)
doi:10.1088/1475-7516/2012/04/010
[arXiv:1202.2962 [hep-ph]].

\bibitem{Yaguna:2019cvp}
C.~E.~Yaguna and \'O.~Zapata,
JHEP \textbf{03}, 109 (2020)
doi:10.1007/JHEP03(2020)109
[arXiv:1911.05515 [hep-ph]].

\bibitem{Belanger:2020hyh}
G.~B\'elanger, A.~Pukhov, C.~E.~Yaguna and \'O.~Zapata,
JHEP \textbf{09}, 030 (2020)
doi:10.1007/JHEP09(2020)030
[arXiv:2006.14922 [hep-ph]].

\bibitem{Aranda:2019vda}
A.~Aranda, D.~Hern\'andez-Otero, J.~Hern\'andez-Sanchez, V.~Keus, S.~Moretti, D.~Rojas-Ciofalo and T.~Shindou,
Phys. Rev. D \textbf{103}, no.1, 015023 (2021)
doi:10.1103/PhysRevD.103.015023
[arXiv:1907.12470 [hep-ph]].

\bibitem{Keus:2014jha}
V.~Keus, S.~F.~King, S.~Moretti and D.~Sokolowska,
JHEP \textbf{11}, 016 (2014)
doi:10.1007/JHEP11(2014)016
[arXiv:1407.7859 [hep-ph]].

\bibitem{Keus:2014isa}
V.~Keus, S.~F.~King and S.~Moretti,
Phys. Rev. D \textbf{90}, no.7, 075015 (2014)
doi:10.1103/PhysRevD.90.075015
[arXiv:1408.0796 [hep-ph]].

\bibitem{Keus:2015xya}
V.~Keus, S.~F.~King, S.~Moretti and D.~Sokolowska,
JHEP \textbf{11}, 003 (2015)
doi:10.1007/JHEP11(2015)003
[arXiv:1507.08433 [hep-ph]].

\bibitem{Cordero-Cid:2016krd}
A.~Cordero-Cid, J.~Hern\'andez-S\'anchez, V.~Keus, S.~F.~King, S.~Moretti, D.~Rojas and D.~Soko\l{}owska,
JHEP \textbf{12}, 014 (2016)
doi:10.1007/JHEP12(2016)014
[arXiv:1608.01673 [hep-ph]].

\bibitem{Cordero:2017owj}
A.~Cordero, J.~Hernandez-Sanchez, V.~Keus, S.~F.~King, S.~Moretti, D.~Rojas and D.~Sokolowska,
JHEP \textbf{05}, 030 (2018)
doi:10.1007/JHEP05(2018)030
[arXiv:1712.09598 [hep-ph]].

\bibitem{Cordero-Cid:2018man}
A.~Cordero-Cid, J.~Hern\'andez-S\'anchez, V.~Keus, S.~Moretti, D.~Rojas and D.~Soko\l{}owska,
Eur. Phys. J. C \textbf{80}, no.2, 135 (2020)
doi:10.1140/epjc/s10052-020-7689-0
[arXiv:1812.00820 [hep-ph]].

\bibitem{Keus:2019szx}
V.~Keus,
Phys. Rev. D \textbf{101}, no.7, 073007 (2020)
doi:10.1103/PhysRevD.101.073007
[arXiv:1909.09234 [hep-ph]].

\bibitem{Cordero-Cid:2020yba}
A.~Cordero-Cid, J.~Hern\'andez-S\'anchez, V.~Keus, S.~Moretti, D.~Rojas-Ciofalo and D.~Soko\l{}owska,
Phys. Rev. D \textbf{101}, no.9, 095023 (2020)
doi:10.1103/PhysRevD.101.095023
[arXiv:2002.04616 [hep-ph]].

\bibitem{Keus:2021dti}
V.~Keus and K.~Tuominen,
Phys. Rev. D \textbf{104}, no.6, 063533 (2021)
doi:10.1103/PhysRevD.104.063533
[arXiv:2102.07777 [hep-ph]].

\bibitem{Weinberg:1976hu}
S.~Weinberg,
Phys. Rev. Lett. \textbf{37}, 657 (1976)
doi:10.1103/PhysRevLett.37.657

\bibitem{Ivanov:2012fp}
I.~P.~Ivanov and E.~Vdovin,
Eur. Phys. J. C \textbf{73}, no.2, 2309 (2013)
doi:10.1140/epjc/s10052-013-2309-x
[arXiv:1210.6553 [hep-ph]].

\bibitem{Keus:2013hya}
V.~Keus, S.~F.~King and S.~Moretti,
JHEP \textbf{01}, 052 (2014)
doi:10.1007/JHEP01(2014)052
[arXiv:1310.8253 [hep-ph]].

\bibitem{Arina:2009um}
C.~Arina, F.~S.~Ling and M.~H.~G.~Tytgat,
JCAP \textbf{10}, 018 (2009)
doi:10.1088/1475-7516/2009/10/018
[arXiv:0907.0430 [hep-ph]].

\bibitem{Nezri:2009jd}
E.~Nezri, M.~H.~G.~Tytgat and G.~Vertongen,
JCAP \textbf{04}, 014 (2009)
doi:10.1088/1475-7516/2009/04/014
[arXiv:0901.2556 [hep-ph]].

\bibitem{Miao:2010rg}
X.~Miao, S.~Su and B.~Thomas,
Phys. Rev. D \textbf{82}, 035009 (2010)
doi:10.1103/PhysRevD.82.035009
[arXiv:1005.0090 [hep-ph]].

\bibitem{Gustafsson:2012aj}
M.~Gustafsson, S.~Rydbeck, L.~Lopez-Honorez and E.~Lundstrom,
Phys. Rev. D \textbf{86}, 075019 (2012)
doi:10.1103/PhysRevD.86.075019
[arXiv:1206.6316 [hep-ph]].

\bibitem{Arhrib:2012ia}
A.~Arhrib, R.~Benbrik and N.~Gaur,
Phys. Rev. D \textbf{85}, 095021 (2012)
doi:10.1103/PhysRevD.85.095021
[arXiv:1201.2644 [hep-ph]].

\bibitem{Krawczyk:2013pea}
M.~Krawczyk, D.~Soko\l{}owska, P.~Swaczyna and B.~\'Swie\.zewska,
Acta Phys. Polon. B \textbf{44}, no.11, 2163-2170 (2013)
doi:10.5506/APhysPolB.44.2163
[arXiv:1309.7880 [hep-ph]].

\bibitem{Goudelis:2013uca}
A.~Goudelis, B.~Herrmann and O.~St\r{a}l,
JHEP \textbf{09}, 106 (2013)
doi:10.1007/JHEP09(2013)106
[arXiv:1303.3010 [hep-ph]].

\bibitem{Arhrib:2013ela}
A.~Arhrib, Y.~L.~S.~Tsai, Q.~Yuan and T.~C.~Yuan,
JCAP \textbf{06}, 030 (2014)
doi:10.1088/1475-7516/2014/06/030
[arXiv:1310.0358 [hep-ph]].

\bibitem{Krawczyk:2015vka}
M.~Krawczyk, M.~Matej, D.~Soko\l{}owska and B.~\'Swie\.zewska,
Acta Phys. Polon. B \textbf{46}, no.1, 169-179 (2015)
doi:10.5506/APhysPolB.46.169
[arXiv:1501.04529 [hep-ph]].

\bibitem{Ilnicka:2015jba}
A.~Ilnicka, M.~Krawczyk and T.~Robens,
Phys. Rev. D \textbf{93}, no.5, 055026 (2016)
doi:10.1103/PhysRevD.93.055026
[arXiv:1508.01671 [hep-ph]].

\bibitem{Diaz:2015pyv}
M.~A.~D\'\i{}az, B.~Koch and S.~Urrutia-Quiroga,
Adv. High Energy Phys. \textbf{2016}, 8278375 (2016)
doi:10.1155/2016/8278375
[arXiv:1511.04429 [hep-ph]].

\bibitem{Modak:2015uda}
K.~P.~Modak and D.~Majumdar,
Astrophys. J. Suppl. \textbf{219}, no.2, 37 (2015)
doi:10.1088/0067-0049/219/2/37
[arXiv:1502.05682 [hep-ph]].

\bibitem{Queiroz:2015utg}
F.~S.~Queiroz and C.~E.~Yaguna,
JCAP \textbf{02}, 038 (2016)
doi:10.1088/1475-7516/2016/02/038
[arXiv:1511.05967 [hep-ph]].

\bibitem{Garcia-Cely:2015khw}
C.~Garcia-Cely, M.~Gustafsson and A.~Ibarra,
JCAP \textbf{02}, 043 (2016)
doi:10.1088/1475-7516/2016/02/043
[arXiv:1512.02801 [hep-ph]].

\bibitem{Hashemi:2016wup}
M.~Hashemi and S.~Najjari,
Eur. Phys. J. C \textbf{77}, no.9, 592 (2017)
doi:10.1140/epjc/s10052-017-5159-0
[arXiv:1611.07827 [hep-ph]].

\bibitem{Poulose:2016lvz}
P.~Poulose, S.~Sahoo and K.~Sridhar,
Phys. Lett. B \textbf{765}, 300-306 (2017)
doi:10.1016/j.physletb.2016.12.022
[arXiv:1604.03045 [hep-ph]].

\bibitem{Alves:2016bib}
A.~Alves, D.~A.~Camargo, A.~G.~Dias, R.~Longas, C.~C.~Nishi and F.~S.~Queiroz,
JHEP \textbf{10}, 015 (2016)
doi:10.1007/JHEP10(2016)015
[arXiv:1606.07086 [hep-ph]].

\bibitem{Datta:2016nfz}
A.~Datta, N.~Ganguly, N.~Khan and S.~Rakshit,
Phys. Rev. D \textbf{95}, no.1, 015017 (2017)
doi:10.1103/PhysRevD.95.015017
[arXiv:1610.00648 [hep-ph]].

\bibitem{Belyaev:2016lok}
A.~Belyaev, G.~Cacciapaglia, I.~P.~Ivanov, F.~Rojas-Abatte and M.~Thomas,
Phys. Rev. D \textbf{97}, no.3, 035011 (2018)
doi:10.1103/PhysRevD.97.035011
[arXiv:1612.00511 [hep-ph]].

\bibitem{Belyaev:2018ext}
A.~Belyaev, T.~R.~Fernandez Perez Tomei, P.~G.~Mercadante, C.~S.~Moon, S.~Moretti, S.~F.~Novaes, L.~Panizzi, F.~Rojas and M.~Thomas,
Phys. Rev. D \textbf{99}, no.1, 015011 (2019)
doi:10.1103/PhysRevD.99.015011
[arXiv:1809.00933 [hep-ph]].

\bibitem{Sokolowska:2019xhe}
D.~Sokolowska, J.~Kalinowski, J.~Klamka, P.~Sopicki, A.~F.~Zarnecki, W.~Kotlarski and T.~Robens,
PoS \textbf{EPS-HEP2019}, 570 (2020)
doi:10.22323/1.364.0570
[arXiv:1911.06254 [hep-ph]].

\bibitem{Kalinowski:2019cxe}
J.~Kalinowski, W.~Kotlarski, T.~Robens, D.~Sokolowska and A.~F.~Zarnecki,
J. Phys. Conf. Ser. \textbf{1586}, no.1, 012023 (2020)
doi:10.1088/1742-6596/1586/1/012023
[arXiv:1903.04456 [hep-ph]].

\bibitem{Hernandez-Sanchez:2020aop}
J.~Hernandez-Sanchez, V.~Keus, S.~Moretti, D.~Rojas-Ciofalo and D.~Sokolowska,
[arXiv:2012.11621 [hep-ph]].

\bibitem{Hernandez-Sanchez:2022dnn}
J.~Hernandez-Sanchez, V.~Keus, S.~Moretti and D.~Sokolowska,
JHEP \textbf{03}, 045 (2023)
doi:10.1007/JHEP03(2023)045
[arXiv:2202.10514 [hep-ph]].

\bibitem{Khalil:2020syr}
S.~Khalil, S.~Moretti, D.~Rojas-Ciofalo and H.~Waltari,
Phys. Rev. D \textbf{102}, no.7, 075039 (2020)
doi:10.1103/PhysRevD.102.075039
[arXiv:2007.10966 [hep-ph]].

\bibitem{Ivanov:2011ae}
I.~P.~Ivanov, V.~Keus and E.~Vdovin,
J. Phys. A \textbf{45}, 215201 (2012)
doi:10.1088/1751-8113/45/21/215201
[arXiv:1112.1660 [math-ph]].

\bibitem{Keus:2016orl}
V.~Keus,
PoS \textbf{CHARGED2016}, 017 (2016)
doi:10.22323/1.286.0017
[arXiv:1612.03629 [hep-ph]].

\bibitem{Belanger:2006is}
G.~Belanger, F.~Boudjema, A.~Pukhov and A.~Semenov,
Comput. Phys. Commun. \textbf{176}, 367-382 (2007)
doi:10.1016/j.cpc.2006.11.008
[arXiv:hep-ph/0607059 [hep-ph]].

\bibitem{Alwall:2014hca}
J.~Alwall, R.~Frederix, S.~Frixione, V.~Hirschi, F.~Maltoni, O.~Mattelaer, H.~S.~Shao, T.~Stelzer, P.~Torrielli and M.~Zaro,
JHEP \textbf{07}, 079 (2014)
doi:10.1007/JHEP07(2014)079
[arXiv:1405.0301 [hep-ph]].

\bibitem{deFavereau:2013fsa}
J.~de Favereau \textit{et al.} [DELPHES 3],
JHEP \textbf{02}, 057 (2014)
doi:10.1007/JHEP02(2014)057
[arXiv:1307.6346 [hep-ex]].

\bibitem{Sjostrand:2006za}
T.~Sjostrand, S.~Mrenna and P.~Z.~Skands,
JHEP \textbf{05}, 026 (2006)
doi:10.1088/1126-6708/2006/05/026
[arXiv:hep-ph/0603175 [hep-ph]].

\bibitem{Grzadkowski:2009bt}
B.~Grzadkowski, O.~M.~Ogreid and P.~Osland,
Phys. Rev. D \textbf{80}, 055013 (2009)
doi:10.1103/PhysRevD.80.055013
[arXiv:0904.2173 [hep-ph]].

\bibitem{Faro:2019vcd}
F.~S.~Faro and I.~P.~Ivanov,
Phys. Rev. D \textbf{100}, no.3, 035038 (2019)
doi:10.1103/PhysRevD.100.035038
[arXiv:1907.01963 [hep-ph]].

\bibitem{Baak:2014ora}
M.~Baak \textit{et al.} [Gfitter Group],
Eur. Phys. J. C \textbf{74}, 3046 (2014)
doi:10.1140/epjc/s10052-014-3046-5
[arXiv:1407.3792 [hep-ph]].

\bibitem{Lundstrom:2008ai}
E.~Lundstrom, M.~Gustafsson and J.~Edsjo,
Phys. Rev. D \textbf{79}, 035013 (2009)
doi:10.1103/PhysRevD.79.035013
[arXiv:0810.3924 [hep-ph]].

\bibitem{Pierce:2007ut}
A.~Pierce and J.~Thaler,
JHEP \textbf{08}, 026 (2007)
doi:10.1088/1126-6708/2007/08/026
[arXiv:hep-ph/0703056 [hep-ph]].

\bibitem{Heisig:2018kfq}
J.~Heisig, S.~Kraml and A.~Lessa,
Phys. Lett. B \textbf{788}, 87-95 (2019)
doi:10.1016/j.physletb.2018.10.049
[arXiv:1808.05229 [hep-ph]].

\bibitem{ATLAS:2015yey}
G.~Aad \textit{et al.} [ATLAS and CMS],
Phys. Rev. Lett. \textbf{114}, 191803 (2015)
doi:10.1103/PhysRevLett.114.191803
[arXiv:1503.07589 [hep-ex]].

\bibitem{Sirunyan:2019twz}
A.~M.~Sirunyan \textit{et al.} [CMS],
Phys. Rev. D \textbf{99}, no.11, 112003 (2019)
doi:10.1103/PhysRevD.99.112003
[arXiv:1901.00174 [hep-ex]].

\bibitem{Khachatryan:2016vau}
G.~Aad \textit{et al.} [ATLAS and CMS],
JHEP \textbf{08}, 045 (2016)
doi:10.1007/JHEP08(2016)045
[arXiv:1606.02266 [hep-ex]].

\bibitem{Sirunyan:2018owy}
A.~M.~Sirunyan \textit{et al.} [CMS],
Phys. Lett. B \textbf{793}, 520-551 (2019)
doi:10.1016/j.physletb.2019.04.025
[arXiv:1809.05937 [hep-ex]].

\bibitem{Aaboud:2019rtt}
M.~Aaboud \textit{et al.} [ATLAS],
Phys. Rev. Lett. \textbf{122}, no.23, 231801 (2019)
doi:10.1103/PhysRevLett.122.231801
[arXiv:1904.05105 [hep-ex]].

\bibitem{Aghanim:2018eyx}
N.~Aghanim \textit{et al.} [Planck],
Astron. Astrophys. \textbf{641}, A6 (2020)
[erratum: Astron. Astrophys. \textbf{652}, C4 (2021)]
doi:10.1051/0004-6361/201833910
[arXiv:1807.06209 [astro-ph.CO]].

\bibitem{Aprile:2018dbl}
E.~Aprile \textit{et al.} [XENON],
Phys. Rev. Lett. \textbf{121}, no.11, 111302 (2018)
doi:10.1103/PhysRevLett.121.111302
[arXiv:1805.12562 [astro-ph.CO]].

\bibitem{PandaX-4T:2021bab}
Y.~Meng \textit{et al.} [PandaX-4T],
Phys. Rev. Lett. \textbf{127}, no.26, 261802 (2021)
doi:10.1103/PhysRevLett.127.261802
[arXiv:2107.13438 [hep-ex]].

\bibitem{Ackermann:2015zua}
M.~Ackermann \textit{et al.} [Fermi-LAT],
Phys. Rev. Lett. \textbf{115}, no.23, 231301 (2015)
doi:10.1103/PhysRevLett.115.231301
[arXiv:1503.02641 [astro-ph.HE]].

\bibitem{Cirelli:2013hv}
M.~Cirelli and G.~Giesen,
JCAP \textbf{04}, 015 (2013)
doi:10.1088/1475-7516/2013/04/015
[arXiv:1301.7079 [hep-ph]].

\bibitem{Behnke:2013xla}
T.~Behnke, J.~E.~Brau, B.~Foster, J.~Fuster, M.~Harrison, J.~M.~Paterson, M.~Peskin, M.~Stanitzki, N.~Walker and H.~Yamamoto,
[arXiv:1306.6327 [physics.acc-ph]].

\bibitem{ILC:2013jhg}
H.~Baer \textit{et al.} [ILC],
[arXiv:1306.6352 [hep-ph]].

\bibitem{Adolphsen:2013jya}
C.~Adolphsen, M.~Barone, B.~Barish, K.~Buesser, P.~Burrows, J.~Carwardine, J.~Clark, H.~Mainaud Durand, G.~Dugan and E.~Elsen, \textit{et al.}
[arXiv:1306.6353 [physics.acc-ph]].

\bibitem{Adolphsen:2013kya}
C.~Adolphsen, M.~Barone, B.~Barish, K.~Buesser, P.~Burrows, J.~Carwardine, J.~Clark, H.~Mainaud Durand, G.~Dugan and E.~Elsen, \textit{et al.}
[arXiv:1306.6328 [physics.acc-ph]].

\bibitem{Behnke:2013lya}
T.~Behnke, J.~E.~Brau, P.~N.~Burrows, J.~Fuster, M.~Peskin, M.~Stanitzki, Y.~Sugimoto, S.~Yamada, H.~Yamamoto and H.~Abramowicz, \textit{et al.}
[arXiv:1306.6329 [physics.ins-det]].

\end{thebibliography}
\end{document}